\documentclass[a4paper,11pt,notoc]{article}
\usepackage{jheppub}
\usepackage{float}
\usepackage{subfig}
\usepackage[table]{xcolor}
\usepackage[compat=1.1.0]{tikz-feynman}
\usepackage{feynmp}
\usepackage{color}
\usepackage{cleveref}

\crefname{section}{Sec.}{Secs.}
\crefname{table}{Tab.}{Tabs.}
\crefname{figure}{Fig.}{Figs.}
\crefname{equation}{Eq.}{Eqs.}
\crefname{appendix}{Appendix}{Appendices}

\newcommand{\be}{\begin{equation}}
\newcommand{\ee}{\end{equation}}
\def\bsp#1\esp{\begin{split}#1\end{split}}
\def\bpm{\begin{pmatrix}}
\def\epm{\end{pmatrix}}

\def\gsim{\raise0.3ex\hbox{$\;>$\kern-0.75em\raise-1.1ex\hbox{$\sim\;$}}}

\preprint{LAPTH-060/22}

\title{Accommodating muon $\boldsymbol{(g-2)}$ and leptogenesis in a scotogenic model}

\author[a]{A.\ Alvarez,}
\author[a]{~A.\ Banik,}
\author[a]{~R.\ Cepedello,}
\author[b]{~B.\ Herrmann,}
\author[a]{~W.\ Porod,}
\author[b]{~M.\ Sarazin,}
\author[a]{~M.\ Schnelke}

\affiliation[a]{Institut f\"ur Theoretische Physik und Astrophysik, Campus Hubland Nord, Univ.\ W\"urzburg, D-97074 W\"urzburg, Germany}
\affiliation[b]{LAPTh, Univ.\ Savoie Mont Blanc, CNRS, F-74000 Annecy, France}

\emailAdd{alexandre.alvarez@physik.uni-wuerzburg.de}
\emailAdd{amitayus.banik@physik.uni-wuerzburg.de}
\emailAdd{ricardo.cepedello@physik.uni-wuerzburg.de}
\emailAdd{herrmann@lapth.cnrs.fr}
\emailAdd{porod@physik.uni-wuerzburg.de}
\emailAdd{sarazin@lapth.cnrs.fr}
\emailAdd{moritz.schnelke@stud-mail.uni-wuerzburg.de}

\abstract{We present a detailed study of a scotogenic model accommodating dark matter, neutrino masses and the anomalous magnetic moment of the muon while being consistent with the existing constraints on flavour violating decays of the leptons. Moreover, this model offers the possibility to explain the baryon asymmetry of the Universe via leptogenesis. We determine the viable regions of the model's parameter space in view of dark matter and flavour constraints using a Markov Chain Monte Carlo setup combined with a particular procedure to accommodate neutrino masses and the anomalous magnetic moment of the muon at the same time. We also discuss briefly the resulting collider phenomenology.
}


\makeatletter
\gdef\@fpheader{}
\vspace*{0cm}
\makeatother

\begin{document}
\notoc
\maketitle
\flushbottom

% ===========================================================================
\section{Introduction}
\label{Sec:Introduction}
% ======================================================

The Standard Model (SM)
gives an accurate description for most of the data up to the TeV scale.
Despite its successes, it should nevertheless be considered as an effective theory, which has to be embedded in a more fundamental framework. One reason is the flavour hierarchies in the fermion sector for which we do not know the underlying principle governing its structures. Moreover, there are several experimental observations which require an extension of the SM. This includes neutrinos, which are massless in the SM, but need to be massive in view of neutrino oscillations experiments \cite{deSalas:2020pgw}. Strong arguments from cosmology underline the call for new physics beyond the Standard Model (BSM), such as the presence of dark matter (DM) \cite{Planck}, as well as the baryon asymmetry observed in the Universe \cite{Planck}. The SM is also challenged by precision measurements of the anomalous magnetic moment of the muon \cite{Muong-2:2006rrc, Aoyama:2020ynm, Muong-2:2021ojo}.

Most of these deviations are located in the lepton sector. Moreover, despite the fact that it ultimately concerns the hadronic sector, the baryon asymmetry can be explained through leptogenesis \cite{Fukugita:1986hr,Buchmuller:2004nz,Davidson:2008bu}, a mechanism stemming from the leptonic sector and translating the generated lepton asymmetry to the hadronic sector through the sphaleron processes. Finally, it is worth noting that generating neutrino masses generally leads to the opening of lepton flavour violating processes, involving, for example, transitions from electronic to muonic states, which are strongly constrained by very precise experimental data. We take this as a motivation to consider models featuring new contributions to the lepton sector, while the hadronic sector may be less relevant to explain the above shortcomings of the SM.

One potential class of such frameworks are the so-called scotogenic models, originally aiming at the simultaneous explanation of neutrino masses and cold dark matter. The two are linked in the sense that neutrino masses are generated radiatively through particles and couplings from the dark sector. After the first works on minimal scotogenic realisation \cite{Ma:2006km, Toma:2013zsa, Vicente:2014wga, Fraser:2014yha, Baumholzer:2019twf}, more complex models have emerged in recent years, studied mainly at the level of dark matter phenomenology and lepton flavour violating observables, see for example \cite{Toma:2013zsa,Rocha-Moran:2016enp, Avila:2019hhv, Ahriche:2020pwq, DeRomeri:2021yjo, Boruah:2021ayj}. A general classification of viable scotogenic frameworks can be found in Ref.~\cite{Restrepo2013}. Recently, two of us have studied a particular framework, the so-called `T1-2-A' model, where the SM is extended by a scalar doublet, a scalar singlet, a fermionic Dirac doublet, and a fermionic singlet \cite{Esch:2014jpa, Sarazin:2021nwo}. This setup features a very predictive dark matter phenomenology, especially for fermionic dark matter \cite{Sarazin:2021nwo} and, in principle, it can explain the recent measurements of the anomalous magnetic moment of the muon. However, the corresponding region in parameter space is excluded by the the constraints on flavour violating decays of the leptons. Moreover, the `T1-2-A' setup fails to accommodate leptogenesis as an explanation for the observed baryon asymmetry. 

In the present work, we extend the `T1-2-A' setup by adding an additional fermionic singlet. The additional degrees of freedom allow for the successful generation of three non-zero neutrino masses, while the couplings can be chosen such that the deviation related to the anomalous magnetic moment of the muon can be accommodated while being consistent with the bounds on flavour violating lepton decays. Moreover, this set-up allows for an explanation of the observed baryon asymmetry via leptogenesis, as we will demonstrate below. 

Our paper is organised as follows: in \cref{Sec:Model}, we start by introducing the scotogenic model under consideration. \cref{Sec:gminus2} is then devoted to a discussion of the anomalous magnetic moment within our model and the required coupling hierarchies. In \cref{Sec:observables}, we discuss the applied constraints and the observables of our interest. The results from our Markov Chain Monte Carlo (MCMC) analysis are presented in \cref{Sec:results}, where we analyse the parameter space, charged lepton flavour violating decays, dark matter observables and discuss collider-related aspects. In \cref{Sec:Leptogenesis}, we present our findings concerning leptogenesis as a means to generate the baryon asymmetry. Conclusions are drawn in \cref{Sec:Conclusion}.

% ======================================================
\section{Model}
\label{Sec:Model}
% ======================================================

We consider a scotogenic framework extending the SM by two Weyl fermion $SU(2)_L$ doublets, $\Psi_1$ and $\Psi_2$, two Majorana fermion singlets, $F_1$ and $F_2$, a scalar $SU(2)_L$ doublet, $\eta$, and a real scalar singlet, $S$. In addition, we assume a $\mathbb{Z}_2$-symmetry under which the SM fields are even and the additional ones are odd. This ensures neutrino mass generation at the one-loop level together with the existence of a stable dark matter candidate. We note for completeness that the additional fields are singlets with respect to $SU(3)_C$.

The new field content including their respective representations under $SU(2)_L\times U(1)_Y$ is summarised in \cref{Tab:T12AQuantumNumbers}. 
In the following subsections, we briefly summarise the different sectors, present the corresponding Lagrangian, and set the notation.

\begin{table}
    \centering
    \begin{tabular}{|c||c|c|c|c||c|c|}
    \hline
            \     & ~$\Psi_{1}$~ & ~$\Psi_2$~ & ~$F_1$~ & ~$F_2$~ & ~$\eta$~ & ~$S$~ \\
            \hline
            \hline
      $SU(2)_L$   &  $\mathbf{2}$ & $\mathbf{2}$ & $\mathbf{1}$ & $\mathbf{1}$ & $\mathbf{2}$ & $\mathbf{1}$ \\
      \hline
        $U(1)_Y$  & -1 & 1 & 0 & 0 & 1 & 0 \\
        \hline
    \end{tabular}
    \caption{Field content of the scotogenic model under consideration beyond the Standard Model fields.} 
    \label{Tab:T12AQuantumNumbers}
\end{table}

% ======================================================
\subsection{The scalar sector}
\label{Sec:ScalarSector}
% ======================================================

The scalar sector of the model consists of the SM Higgs doublet $H$, an additional real singlet $S$, and a $SU(2)_L$ doublet $\eta$. Their charges are given in \cref{Tab:T12AQuantumNumbers}. Upon electroweak symmetry breaking (EWSB), which involves the Higgs doublet only, the doublets can be expanded into components according to
\begin{align}
	H ~=~ \begin{pmatrix} G^+ \\ \frac{1}{\sqrt{2}} \big[ v + h^0 + i G^0 \big] \end{pmatrix}, \qquad
	\eta ~=~ \begin{pmatrix} \eta^+ \\ \frac{1}{\sqrt{2}} \big[ \eta^0 + i A^0 \big] \end{pmatrix} \,.
\end{align}
Here, $h^0$ is the SM Higgs boson, $G^0$ and $G^+$ are the would-be Goldstone bosons, and $v = \sqrt{2} \langle H \rangle \approx 246$ GeV denotes the vacuum expectation value (VEV). Moreover, $\eta^0$ and $A^0$ are $CP$-even and $CP$-odd neutral scalars, and $\eta^+$ is a charged scalar. Neither $S$ nor $\eta$ may obtain a VEV due to the assumed $\mathbb{Z}_2$-symmetry.

The scalar potential of the model is given by
\begin{align}
    \begin{split}
	V_{\rm scalar} ~=&~ M_H^2 \big| H \big|^2 + \lambda_H \big| H \big|^4 + \frac{1}{2} M_S^2 S^2 + \frac{1}{2} \lambda_{4S} S^4 + M_{\eta}^2 \big| \eta \big|^2 + \lambda_{4\eta} \big| \eta \big|^4 \\ &~+ \frac{1}{2} \lambda_S S^2 \big| H \big|^2  +  \frac{1}{2} \lambda_{S\eta} S^2 \big| \eta \big|^2+\lambda_{\eta} \big| \eta \big|^2 \big| H \big|^2 
		+ \lambda'_{\eta} \big| H \eta^{\dag} \big|^2 \\ &~+ \frac{1}{2} \lambda''_{\eta} \Big[ \big( H \eta^{\dag} \big)^2 + {\rm h.c.}\Big] + \alpha \Big[ S H \eta^{\dag} + {\rm h.c.} \Big] \,.
    \end{split}
	\label{Eq:ScalarPotential}
\end{align}
The first two terms are the SM part related to the Higgs doublet $H$.  We assume here for simplicity that $\lambda''_{\eta}$ and $\alpha$ are real. After EWSB, the usual minimisation relation in the Higgs sector,
\begin{align}
	m^2_{h^0} ~=~ -2 M_H^2 ~=~ 2 \lambda_H v^2 \,,
	\label{Eq:HiggsMass}
\end{align}
allows to eliminate the mass parameter $M_H^2$ in favour of the Higgs self-coupling $\lambda_H$. Imposing $m_{h^0} \approx 125$ GeV leads to a tree-level value of $\lambda_H \approx 0.13$.

The mass matrix of the neutral scalars in the basis $\{ S, \eta^0, A^0 \}$ reads as
\begin{equation}
    {\cal M}_{\phi}^2 ~=~ \begin{pmatrix} M^2_S + \frac{1}{2} v^2 \lambda_S & v \alpha & 0 \\ v \alpha & M^2_{\eta} + \frac{1}{2}v^2 \lambda_L & 0 \\ 0 & 0 & M^2_{\eta} + \frac{1}{2} v^2 \lambda_A \end{pmatrix} \,,
    \label{eq:scalar_mass_matrix}
\end{equation}
after EWSB. Here, we have defined $\lambda_{L,A} ~=~ \lambda_{\eta} + \lambda'_{\eta} \pm \lambda''_{\eta}$. We order the mass eigenstates as follows
\begin{align}    
\big( \phi^0_1, \phi^0_2, A^0 \big)^T ~=~ U_{\phi} \, \big( S, \eta^0, A^0 \big)^T \,.
\label{eq:def_Uphi}
\end{align}
The corresponding squared masses at tree-level read as
\begin{align}
    m^2_{\phi^0_{1,2}} ~&=~ \frac{1}{2} \left[ M^2_S + M^2_{\eta} + \frac{1}{2} v^2 \left( \lambda_S + \lambda_L \right) \mp \sqrt{ \left[ M^2_S - M^2_{\eta} + \frac{1}{2}v^2 \left( \lambda_S - \lambda_L \right) \right]^2 + 4 v^2 \alpha^2} \right] \, , \nonumber \\
    m^2_{A^0} ~&=~ M^2_{\eta} + \frac{1}{2} v^2 \lambda_A \, ,
\end{align}
where $m_{\phi^0_1} < m_{\phi^0_2}$. Finally, the tree-level mass of the charged scalars is given by
\begin{equation}
    m^2_{\eta^{\pm}} ~=~ M^2_{\eta} + \frac{1}{2}v^2 \lambda_{\eta} \,.
\end{equation}

% ======================================================
\subsection{The fermion sector}
\label{Sec:FermionSector}
% ======================================================

The Lagrangian for the additional fermions presented in \cref{Tab:T12AQuantumNumbers} reads
\begin{align}
\begin{split}
	{\cal L}_{\rm fermion} ~=~& i  \Big( \overline{\Psi}_j \sigma^{\mu}D_{\mu} \Psi_j + \frac{1}{2} \overline{F}_j \sigma^{\mu} \partial_{\mu} F_j \Big) 
	- \frac{1}{2} M_{F_{ij}} F_i F_j \\ 
	&  - M_{\Psi} \Psi_1 \Psi_2 - y_{1i} \Psi_1 H F_i - y_{2i} \Psi_2 \tilde H F_i \\
	& - g_{\Psi}^k \Psi_2 L_k S - g_{F_j}^k \eta L_k F_j - g_R^k e^c_k \tilde \eta \Psi_1 + \mathrm{h.c.}
	\label{eqn:fermion_lagrangian}
\end{split}
\end{align}
with $i,j=1,2$ and $k=1,2,3$. $L_k$ and $e^c_k$ denote the left-handed and right-handed leptons, respectively. Moreover, we have introduced the notation $\tilde \phi = i \sigma_2 \phi^*$ for $\phi = H,\eta$. Without loss of generality we work in a basis where $M_{F_{12}} = 0$. Moreover, we impose $|M_1| \le |M_2|$, where we have simplified the notation by setting $M_i = M_{F_{ii}}$ for $i=1,2$. Finally, we adopt the phase-convention $\Psi_1 =(\Psi^0_1, \Psi^-_1)$ and $\Psi_2 =(\Psi^+_2, -\Psi^0_2)$ for the $SU(2)_L$ doublets.

After EWSB, we have a charged heavy Dirac state $\Psi^+$ with mass $M_\Psi$ and four neutral Majorana fermions. Their mass matrix is given in the basis $\{F_1, F_2,\Psi^0_1,\Psi^0_2\}$ as 
\begin{align}
{\cal M}_{\chi^0} ~=~ 
\begin{pmatrix}
M_1 & 0 & \frac{v}{\sqrt{2}} \, y_{11} & \frac{v}{\sqrt{2}} \, y_{21} \\
0 & M_2 & \frac{v}{\sqrt{2}} \, y_{12} & \frac{v}{\sqrt{2}} \, y_{22} \\
\frac{v}{\sqrt{2}} y_{11} & \frac{v}{\sqrt{2}} \, y_{12} & 0 & M_\Psi \\
\frac{v}{\sqrt{2}} y_{21} & \frac{v}{\sqrt{2}} y_{22} & M_\Psi &0 
\end{pmatrix} \,.
\label{eq:fermion_mass_matrix}
\end{align}
This matrix is diagonalised by a unitary matrix $U_\chi$ according to
\begin{align}
    \text{diag}\big(m_{\chi^0_1}, m_{\chi^0_2}, m_{\chi^0_3}, m_{\chi^0_4}\big) ~=~ U_\chi {\cal M}_{\chi^0} U^{-1}_\chi \, ,  
\label{eq:def_Uchi}
\end{align}
with the convention $m_{\chi^0_i} \le m_{\chi^0_j}$ for $i<j$.

%============================================
\subsection{Neutrino masses}
\label{Sec:NuMasses}
%============================================

The main difference with respect to the T1-2-A model discussed in Refs.\ \cite{Esch:2014jpa, Sarazin:2021nwo} is the extra copy of the singlet fermion. Although the mechanism is very similar, in this case, due to the extra degree of freedom, the neutrino mass matrix has rank three instead of two, and consequently, all three active neutrinos will acquire a non-zero mass. After EWSB, rotating to the mass eigenbasis, a Majorana mass term is generated at the one-loop level via the diagram
\begin{equation} 
    \feynmandiagram[small,layered layout,baseline=(d.base),horizontal = f2 to f3] {
f2[particle=\(\nu_i\)] -- [fermion]b[dot]  -- [ edge label'= \(\chi^0_k\)] d[dot] -- [anti fermion] f3[particle=\(\nu_j\)],
b-- [scalar, half left,looseness=1.55, edge label= \(\phi^0_n\)] d ;
}; \quad \equiv \quad \overline{\nu_j^c} \, \big({\cal M}_{\nu} \big)_{ji} \, \nu_i \, ,
\end{equation}
where the neutrino mass matrix can be expressed as
\begin{equation} \label{Eq:mnu}
    {\cal M}_\nu ~=~ \mathcal{G}^T \, M_L \, \mathcal{G} \,.
\end{equation}
This is a well-known structure common to most of the scotogenic models and similar to the type-I seesaw, where the matrix $\mathcal{G}$ contains the couplings defined in Eq.\ \eqref{eqn:fermion_lagrangian} ordered as
\begin{align}
    {\cal G} ~=~ \begin{pmatrix} 
    g_{\Psi}^1 & g_{\Psi}^2 & g_{\Psi}^3 \\
    g_{F_1}^1 & g_{F_1}^2 & g_{F_1}^3 \\
    g_{F_2}^1 & g_{F_2}^2 & g_{F_2}^3 
    \end{pmatrix} \,,
\label{eq:G_matrix}
\end{align}
and $M_L$ is a $3 \times 3$ symmetric matrix which encodes the information of the loop function, and the mixing in the neutral scalar and fermion sectors, defined in Eqs.\ \eqref{eq:def_Uphi} and \eqref{eq:def_Uchi}, respectively. For completeness, we explicitly write the expressions for the components of $M_L$,
\begin{eqnarray} %\label{eq:}
    (M_L)_{11} &=& \sum_{k,n} b_{kn}  (U_{\chi}^{\dagger})^2_{4k} (U_{\phi}^{\dagger})^2_{1n} \, ,
    \\
    (M_L)_{22} &=& \frac{1}{2} \, \sum_{k,n} b_{kn} (U_{\chi}^{\dagger})^2_{1k}  \left[ (U_{\phi}^{\dagger})^2_{2n} - (U_{\phi}^{\dagger})^2_{3n} \right] \, ,
    \\
    (M_L)_{33} &=& \frac{1}{2} \, \sum_{k,n} b_{kn} (U_{\chi}^{\dagger})^2_{2k}  \left[ (U_{\phi}^{\dagger})^2_{2n} - (U_{\phi}^{\dagger})^2_{3n} \right] \, ,
    \\
    (M_L)_{12} = (M_L)_{21} &=& \frac{1}{\sqrt{2}} \, \sum_{k,n} b_{kn} (U_{\chi}^{\dagger})_{1k} (U_{\chi}^{\dagger})_{4k} (U_{\phi}^{\dagger})_{1n} (U_{\phi}^{\dagger})_{2n} \, ,
    \\
    (M_L)_{13} = (M_L)_{31} &=& \frac{1}{\sqrt{2}} \, \sum_{k,n} b_{kn} (U_{\chi}^{\dagger})_{2k} (U_{\chi}^{\dagger})_{4k} (U_{\phi}^{\dagger})_{1n} (U_{\phi}^{\dagger})_{2n} \, ,
    \\
    (M_L)_{23} = (M_L)_{32} &=& \frac{1}{2} \, \sum_{k,n} b_{kn} (U_{\chi}^{\dagger})_{2k} (U_{\chi}^{\dagger})_{1k} \left[ (U_{\phi}^{\dagger})^2_{2n} - (U_{\phi}^{\dagger})^2_{3n} \right] \, ,
\end{eqnarray}
where $k=1,2,3,4$ and $n=1,2,3$. Moreover, the loop integrals are encompassed in the functions
\begin{align}
    b_{kn} ~=~ \frac{1}{16 \pi^2} \frac{m_{\chi^0_k}}{m_{\phi^0_n}^2 - m_{\chi^0_k}^2 } \left[ m_{\chi^0_k}^2  \log m_{\chi^0_k}^2- m_{\phi^0_n}^2 \log m_{\phi^0_n}^2  \right] \,.
\end{align}
We make use of the Casas-Ibarra parametrisation \cite{Casas:2001sr, Basso:2012voo} to express the couplings in \cref{eq:G_matrix} in terms of neutrino oscillation data \cite{deSalas:2020pgw, Gonzalez-Garcia:2021dve} according to
\begin{equation} \label{eq:G_param}
    {\cal G} ~=~ U_L \, D^{-1/2}_L \, R \, D_{\nu}^{1/2} \, U^{*}_{\rm PMNS} \, ,
\end{equation}
where $D_L$ is the diagonal matrix defined by
\begin{align}
   D_L ~=~ U_L^T \, M_L \, U_L \,, 
\end{align}
and $D_{\nu}$ is the diagonal matrix containing the neutrino mass eigenvalues. Finally, $U_{\rm PMNS}$ is the usual unitary matrix relating neutrino flavours to their mass eigenstates, assuming that the charged leptons are already in their mass eigenbasis.

Moreover, as even a precise knowledge of all the parameters and observables in $M_L$ and ${\cal M}_\nu$ does not univocally define ${\cal G}$, the extra degrees of freedom are encoded in the orthogonal $3 \times 3$ matrix $R$. This matrix can be parameterised as
\begin{equation} \label{eq:Rmat}
    R ~=~ \begin{pmatrix}
    c_2 c_3 ~&~ - c_1 s_3 - s_1 s_2 c_3 ~&~ s_1 s_3 - c_1 s_2 c_3 \\
    c_2 s_3 ~&~ c_1 c_3 - s_1 s_2 s_3 ~&~ -s_1 c_3 - c_1 s_2 s_3 \\
    s_2 ~&~ s_1 c_2 ~&~ c_1 c_2
\end{pmatrix} \, ,
\end{equation}
depending on three complex angles $\theta_i$ with $s_i = \sin\theta_i$ and $c_i = \sqrt{1 - s^2_i}$. Note the importance of these degrees of freedom, since they modify the flavour structure of the Yukawa matrix. The latter is of great relevance when considering charged lepton flavour violation and the anomalous magnetic moment, as we will show in the next sections.

% ===========================================================================
\section{The anomalous magnetic moment of the muon}
\label{Sec:gminus2}
% ===========================================================================

As mentioned in the introduction, a deviation persists between the SM prediction and the experimental value of the anomalous magnetic moment of the muon, defined as $a_{\mu} = (g-2)_{\mu}/2$. The discrepancy amounts to a significance of $4.2\sigma$, and leads to the following range for the new physics contribution to $a_{\mu}$ \cite{Aoyama:2020ynm, Muong-2:2021ojo},\footnote{We note here that the SM calculation is currently under discussion due to recent lattice results that weaken the anomaly \cite{Borsanyi:2020mff,Ce:2022kxy,Alexandrou:2022amy}. However, these results are still in tension with $e^+ e^- \to$ hadrons cross-section data and EW precision observables \cite{Keshavarzi:2020bfy, Crivellin:2020zul, Colangelo:2022vok}.}
\begin{align}
    a_{\mu}^{\rm BSM} ~=~ a_{\mu}^{\rm exp} - a_{\mu}^{\rm SM} ~=~ \big( 251 \pm 59 \big) \times 10^{-11} \,.
\end{align}

In general, every scotogenic-like model will contribute to the anomalous magnetic moment of leptons at one-loop level. These contributions can be encoded in the effective electromagnetic (EM) dipole moment operator $c_R^{i j} \, \bar{\ell_i} \sigma_{\mu\nu} P_R \ell_j F^{\mu\nu}$, coming from the operator $\mathcal{O}_{eB} \equiv (\bar L \sigma_{\mu\nu} e_R) H B^{\mu\nu}$ before EWSB \cite{Grzadkowski:2010es}. The diagonal part of the Wilson coefficient $c_R$ is related to $(g-2)$ and the electric dipole moment (EDM), and the off-diagonal part is associated with charged lepton flavour violating (cLFV) processes \cite{Crivellin:2018qmi}. For more details see \cref{Appendix:EMope}.

\begin{figure}
    \centering
    \includegraphics[width=0.4\textwidth]{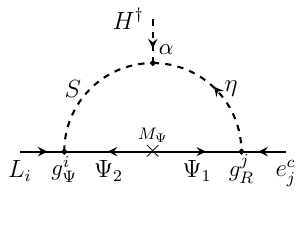} \qquad
    \includegraphics[width=0.4\textwidth]{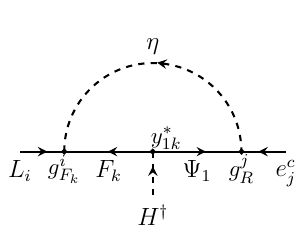}
    \caption{Dominant one-loop contributions to $(g-2)$ and charged LFV processes before EWSB. Arrows indicate the flow of quantum numbers. Couplings are given for clarity, see their explicit definitions in \cref{Sec:Model}. A photon should be attached to the respective charged components.}
    \label{Fig:gm2_diagrams}
\end{figure}

The contribution to $(g-2)_{\mu}$ is generally suppressed by the muon mass. Moreover, as the EM dipole operator connects the left- and right-handed parts of the leptons, while neutrino mass models contain usually only couplings of BSM fields to the left-handed components, such an operator will always be chirally suppressed. Consequently, new physics explanations of $(g-2)_{\mu}$ are pushed towards low mass scales and large non-perturbative couplings. A possible way out is to add new fields outside the neutrino mass mechanism, that couple to $\mu_R$, in order to enhance the contribution to $(g-2)_\mu$ and be able to fit the anomaly within a phenomenologically reasonable parameter space \cite{Arbelaez:2020rbq}. 
Note that this situation is realised in the T1-2-A model \cite{Restrepo2013}, and consequently its extension under consideration here. In both models, there is a coupling $g_R$ of the lepton singlets to $\eta$ and $\Psi_1$, see \cref{eqn:fermion_lagrangian}. The latter two also participate in the generation of the neutrino mass matrix. 
Note that in this way no extra BSM field is needed on top of those involved in the neutrino mass mechanism to have a chirally enhanced contribution to $(g-2)$. The new leading contributions to the anomalous magnetic moment are shown in \cref{Fig:gm2_diagrams}.

We note for completeness, that in the original T1-2-A framework the coupling matrix ${\cal G}$, see \cref{eq:G_matrix}, is a $2\times 3$ matrix, where the relative sizes between the various entries are fixed by the neutrino mixing angles up to one complex angle. An explanation of the muon $(g-2)$ in this model implies large couplings which in turn lead to too large flavour violating decays of the leptons. In our extension of this model, we have more freedom allowing us to circumvent this problem, as we will show in \cref{Sec:clfv}.

Both diagrams depicted in \cref{Fig:gm2_diagrams} also generate a sizeable contribution to strongly constrained LFV processes in the charged sector, in particular $\mu \to e \gamma$ with an upper limit to its branching ratio of $4.2 \times 10^{-13}$ from the MEG collaboration \cite{MEG:2016leq}. Although they seem unavoidable, given that the off-diagonal part of the Yukawa matrix $\mathcal{G}$ is connected to the neutrino mixing (see \cref{eq:G_param}) there are several strategies to get a sizeable contribution to $(g-2)_{\mu}$ while keeping charged LFV under control. For example, cancellations can be found among the several independent contributions to the EM dipole operator. However, such a scenario is not very appealing, as one should reproduce a difference of more than five orders of magnitude between the diagonal and off-diagonal components of the Wilson coefficient $c_R$ \cite{Crivellin:2018qmi}. Another possibility is to assume certain flavour structures for the Yukawa couplings, which suppress the off-diagonal components in favour of the diagonal \cite{Cannoni:2013gq}.

\begin{figure}
    \centering
    \includegraphics[width=0.5\textwidth]{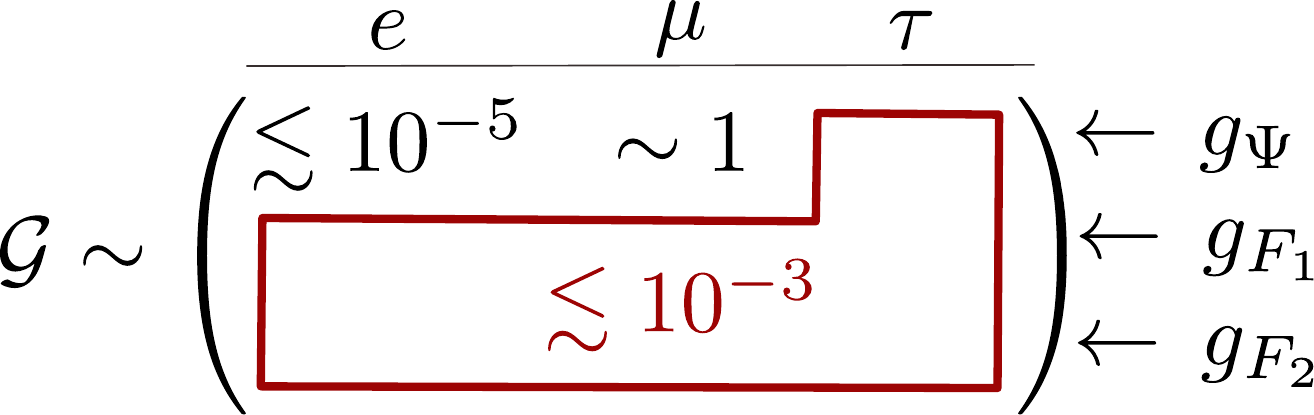}
    \caption{Hierarchy of the Yukawa matrix ${\cal G}$ of \cref{eq:G_matrix} that generates the neutrino masses. This hierarchy is realised for $y_{1,2}$ small and solving for the angles defining the rotation matrix $R$, such that $(g-2)_\mu$ is maximised while charged lepton flavour violating decays are kept under control. See text for more details.}
    \label{Fig:Gmat_order}
\end{figure}

Following the latter approach, we focus for simplicity on a region of the parameter space where the first diagram in \cref{Fig:gm2_diagrams} dominates over the second, as the flavour structure of the diagram is simpler having just two three-component Yukawa vectors involved. We extend the usual Casas-Ibarra parameterisation by the following elements so that these constraints can be easily fulfilled. To do so, we consider $y_{1,2}$ to be small and push the trilinear coupling $\alpha$ to larger values, i.e. we suppress the mixing in the neutral fermion sector, while enhancing the one in the neutral scalar sector. Note that, while $g_R$ is mainly free, $g_F$ and $g_\Psi$ are constrained by the fit to neutrino oscillation data, see Eq.\ \eqref{eq:G_param}. This means that changing $y_{1,2}$ and $\alpha$ not only directly modifies the dominant contributions depicted in \cref{Fig:gm2_diagrams}, but also indirectly suppresses $g_F$ and enhances $g_\Psi$ through the neutrino fit. We are looking for a Yukawa matrix ${\cal G}$ featuring a coupling hierarchy as shown in \cref{Fig:Gmat_order}. Making use of the freedom on the components of $g_R$ as well as on the remaining degrees of freedom in $g_\psi$, stemming from the rotation matrix $R$ appearing in \cref{eq:G_param}, we fit the value of $a_{\mu}^{\rm BSM}$ while keeping the contributions to the lepton flavour violating decays $\mu \to e \gamma$ and $\tau \to \mu \gamma$ under control. 

In practice, for each point of our numerical scan, in the region of the parameter space where $y_{1,2}$ are small, we use the angles of the matrix $R$ given in \eqref{eq:Rmat} to suppress the dominant contribution to cLFV processes while enhancing the diagonal contribution associated to $(g-2)_\mu$\footnote{Actually, solving for two angles is sufficient, such that one angle is left as a free parameter and scanned over for generality.}. Ultimately, we fit the experimental value of the muon $(g-2)$ within its limits by solving for $g_R^2$. With this method, we obtain for each point the correct anomalous magnetic moment, while fulfilling the current limits for charged LFV decays. In this way we can account for the fact that
the neutrino masses, muon $(g-2)$ and the constraint from the cLFV decays pull the parameters in different directions, which requires a certain tuning of the underlying parameters.

We note that this is not the most general approach, as we are selecting a specific region of the parameter space. However, given the complexity of the system, we were not able to find a more general approach that could deliver results within a reasonable computing time. Moreover, the parameter space discussed previously is also preferred for low-scale leptogenesis, as we will show later in the paper.

% ===========================================================================
\section{Constraints and observables}
\label{Sec:observables}
% ===========================================================================

In the spirit of the analysis presented in Ref.\ \cite{Sarazin:2021nwo}, we use an MCMC scan \cite{Markov1971} based on the Metropolis-Hastings algorithm \cite{Metropolis:1953am, Hastings:1970aa} to efficiently scrutinise the parameter space of the model in view of the numerous constraints presented above. This technique, especially powerful for high-dimensional spaces, explores the parameter space iteratively, restricted by a set of constraints through the computation of the likelihood. We refer the reader to Ref.\ \cite{Sarazin:2021nwo} for further details about the implementation of the MCMC.

In addition to implicitly satisfying the constraints from neutrino masses and the anomalous magnetic moment of the muon (see above), we explicitly impose constraints coming from various sectors, comprising dark matter, lepton flavour violating processes, and the mass of the Higgs boson, which is calculated at the one-loop level. 
All constraints are listed in \cref{tab:constraints} together with their associated experimental limits, as well as applied uncertainties applied in our study. Note that for the Higgs mass $m_H$ and the dark matter relic density $\Omega_{\rm CDM}h^2$, the theory uncertainties\footnote{We estimate those on $m_H$ to be of similar size as those in supersymmetric models due to electroweak corrections.} \cite{Allanach:2004rh, Slavich:2020zjv, Boudjema:2011ig, Boudjema:2014gza, Harz:2016dql} are larger than the experimental ones, and, consequently, we apply the theory uncertainties. We also ensure that the lightest $\mathbb{Z}_2$-odd particle, is electrically neutral in order to have a viable, stable DM candidate and to avoid stable charged relics, essentially excluded in the mass range of $\left[1, 10^5\right]$ GeV \cite{ParticleDataGroup:2022pth, Hemmick:1989ns, Kudo:2001ie, Taoso:2007qk}.

\begin{table}
    \centering
    \begin{tabular}{|c|c|}
        \hline
         \textbf{Observable} & \textbf{Constraint}
         \\
         \hline\hline
        $m_H$ & 125.25 $\pm$ 1.0 GeV
        \\
        \hline
        $\Omega_{\rm CDM}h^2$ & 0.120 $\pm$ 0.012
        \\
        \hline
        BR( $\mu^-$ $\to$ $e^- \gamma$ ) & $< 4.2 \times 10^{-13}$
        \\
        \hline
        BR( $\tau^-$ $\to$ $e^- \gamma$ ) & $< 3.3 \times 10^{-8}$
        \\
        \hline
        BR( $\tau^-$ $\to$ $\mu^- \gamma$ ) & $< 4.2 \times 10^{-8}$
        \\
        \hline
        BR( $\mu^-$ $\to$ $e^- e^+ e^-$ ) & $< 1.0 \times 10^{-12}$
        \\
        \hline
        BR( $\tau^-$ $\to$ $e^- e^+ e^-$ ) & $< 2.7 \times 10^{-8}$
        \\
        \hline
        BR( $\tau^-$ $\to$ $\mu^- \mu^+ \mu^-$ ) & $< 2.1 \times 10^{-8}$
        \\
        \hline
        BR( $\tau^-$ $\to$ $e^- \mu^+ \mu^-$ ) & $< 2.7 \times 10^{-8}$
        \\
        \hline
        BR( $\tau^-$ $\to$ $ \mu^- e^+ e^-$ ) & $< 1.8 \times 10^{-8}$
        \\
        \hline
        BR( $\tau^-$ $\to$ $ \mu^- e^+ \mu^-$ ) & $< 1.7 \times 10^{-8}$
        \\
        \hline
        BR( $\tau^-$ $\to$ $ \mu^+ e^- e^-$ ) & $< 1.5 \times 10^{-8}$
        \\
        \hline
    \end{tabular}
    \qquad
    \begin{tabular}{|c|c|}
        \hline
         \textbf{Observable} & \textbf{Constraint}
         \\
         \hline\hline
        BR( $\tau^-$ $\to$ $  e^- \pi $ ) & $< 8.0 \times 10^{-8}$
        \\
        \hline
        BR( $\tau^-$ $\to$ $ e^- \eta$ ) & $< 9.2 \times 10^{-8}$
        \\
        \hline
        BR( $\tau^-$ $\to$ $ e^- \eta'$ ) & $< 1.6 \times 10^{-7}$
        \\
        \hline
        BR( $\tau^-$ $\to$ $  \mu^- \pi $ ) & $< 1.1 \times 10^{-7}$
        \\
        \hline
        BR( $\tau^-$ $\to$ $ \mu^- \eta$ ) & $< 6.5 \times 10^{-8}$
        \\
        \hline
        BR( $\tau^-$ $\to$ $ \mu^- \eta'$ ) & $< 1.3 \times 10^{-7}$
        \\
        \hline
        CR$_{\mu \to e}$(Ti) & $< 4.3 \times 10^{-12}$
        \\
        \hline
        CR$_{\mu \to e}$(Pb) & $< 4.3 \times 10^{-11}$
        \\
        \hline
        CR$_{\mu \to e}$(Au) & $< 7.0 \times 10^{-13}$
        \\
        \hline
        BR( $Z^0$ $\to$ $ e^\pm \mu^\mp  $ ) & $< 7.5 \times 10^{-7}$
        \\
        \hline
        BR( $Z^0$ $\to$ $ e^\pm \tau^\mp $ ) & $< 5.0\times 10^{-6}$
        \\
        \hline
        BR( $Z^0$ $\to$ $ \mu^\pm \tau^\mp$ ) & $< 6.5 \times 10^{-6}$
        \\
        \hline
    \end{tabular}
    \caption{Constraints considered in the MCMC analysis: Higgs mass and cLFV observables \cite{ParticleDataGroup:2022pth} and DM relic density \cite{Planck}. The limits from LUX-ZEPLIN (LZ) \cite{LZ:2022ufs} to the direct detection cross-section are also taken into account. Note that the errors given for $m_H$ and $\Omega_{\rm CDM}h^2$ are not the experimental uncertainties but estimates of the theoretical ones, see text for details. We implement $1\sigma$ intervals using a Gaussian function and $90\%$ C.L. for the limits via a single-sided Gaussian, allowing for a $10\%$ uncertainty.}
    \label{tab:constraints}
\end{table}

In total, our MCMC scan runs over 20 free parameters: Eight couplings in the scalar potential, six Yukawa couplings, five masses, the lightest neutrino mass, and the unconstrained angle of the rotation matrix $R$, which is assumed to be real. The ranges of the scalar and fermion mass parameters are chosen such, that they could be in principle in the reach of high luminosity LHC. The exception is those for the singlet fermions for which we allow a larger range. The reason is that this model can also explain the baryon asymmetry of the Universe via the leptogenesis mechanism as discussed in \cref{Sec:Leptogenesis}. The sign of the quartic couplings $\lambda_H$, $\lambda_{4S}$ and  $\lambda_{4\eta}$ is fixed from the requirement that the scalar potential is bounded from below. We vary all parameters on a logarithmic scale
and assign possible signs on a random basis.

The scan is performed over the parameter ranges specified in \cref{tab:constraints} with 75 chains of 200 points each. The first 35 points of each chain have been deleted in order to keep only the points for which the chains were already well initialised, i.e.\ presenting a phenomenologically viable likelihood value. For the scan we implemented the model in \verb|SARAH-4.14.| \cite{SARAH2014} and generate code for \verb|SPheno-4.0.4| \cite{SPheno2012}, \verb|FlavorKit| \cite{Porod:2014xia} and \verb|micrOMEGAS-5.2.7| \cite{MO2018}. The former two compute the mass spectrum and low energy observables, while the latter evaluates the DM relic density and the direct detection (DD) cross-sections. 

Assuming a Gaussian likelihood of uncorrelated observables, the likelihood associated with a given parameter point $n$ is computed as
\begin{equation}
    {\cal L}_n ~=~ \prod_i {\cal L}^n_i \,,
\end{equation}
where the product runs over the imposed constraints and individual likelihood value ${\cal L}^n_i$ associated to each constraint. In the case of a two-sided limit, i.e.\ for the Higgs mass $m_H$ and the DM relic density $\Omega_{\rm CDM}h^2$, the likelihood is computed following
\begin{equation}
    \ln{\cal L}_n^i ~=~ -\frac{\big( O^n_i - O_i^{\rm exp} \big)^2}{2\sigma^2_i} \,,
    \label{eq:likelihood}
\end{equation}
where $O^n_i$ is the calculated value of the considered observable for the parameter point $n$, $O_i^{\rm exp}$ is the associated experimental value given in \cref{tab:constraints}, and $\sigma_i$ is the associated uncertainty. The likelihood computation for upper limits is implemented as a step function, which is smeared as a single-sided Gaussian with a width of 10\% of the value corresponding to the experimental upper limit. In this case, we have ${\cal L}^n_i = 1$ if the predicted value $O_i^n$ is below the upper limit $O_i^{\rm exp}$. In the opposite case, ${\cal L}^n_i$ is computed according to \cref{eq:likelihood} with $O_i^{\rm exp}$ being the upper limit and $\sigma_i = 0.1 \, O_i^{\rm exp}$.

\begin{table}
    \centering
    \begin{tabular}{|c|c|}
        \hline
         \textbf{Parameter} & \textbf{Interval}
         \\
         \hline\hline
        $\lambda_H$ & [0.1; 0.4]
        \\
        \hline
        $\lambda_{4S}$, $\lambda_{4\eta}$ & [$10^{-7}$; 1]
        \\
        \hline
         %$\lambda_{4\eta}$ & [$10^{-7}$; 1]
        %\\
        %\hline
        $\lambda_{S\eta}$, $\lambda_{S}$ & [-1; 1]
        \\
        \hline
        %$\lambda_{S}$ & [-1; 1]
        %\\
        %\hline
        $\lambda_{\eta}$, $\lambda_{\eta}'$, $\lambda_{\eta}''$ & [-1; 1]
        \\
        \hline
        %$\lambda_{\eta}'$ & [-1; 1]
        %\\
        %\hline
        %$\lambda_{\eta}''$ & [-1; 1]
        %\\
        %\hline
        $\alpha$ & [$-10^4$; $10^4$]
        \\
        \hline
    \end{tabular}
    \qquad
    \begin{tabular}{|c|c|}
        \hline
         \textbf{Parameter} & \textbf{Interval}
         \\
         \hline\hline
         $M_S^2$, $M_\eta^2$ & [$5\times 10^5$; $5\times 10^6$] 
        \\
        \hline
         %$M_\eta^2$ & [$5\cdot 10^5$; $5\cdot 10^6$] 
        %\\
        %\hline
        $M_{1}$, $M_{2}$ & [100; $2\times 10^4$]
        \\
        \hline
        %$M_{F_2}$ & [100; $2\cdot 10^4$] 
        %\\
        %\hline
        $M_{\Psi}$ & [700; 2000] 
        \\
        \hline
        $y_{11}$, $y_{12}$, $y_{21}$, $y_{22}$ & [$-10^{-4}; 10^{-4}$]
        \\
        \hline
        %$y_1^2$ & [$10^{-12}; 10^{-4}$]
        %\\
        %\hline
        %$y_2^1$ & [$10^{-12}; 10^{-4}$]
        %\\
        %\hline
        %$y_2^2$ & [$10^{-12}; 10^{-4}$]
        %\\
        %s\hline
        $m_{\nu_1}$ & [$10^{-32}; 10^{-10}$] 
        \\
        \hline
    \end{tabular}
    \caption{Input parameter for the MCMC scan. All dimensionful quantities are given in GeV.}
    \label{tab:parameters}
\end{table}

% ===========================================================================
\section{Results}
\label{Sec:results}
% ===========================================================================

In this section, we present the main outcome of our MCMC analysis. We shall first show the resulting parameter space for the couplings and then discuss certain observables of interest. In addition, we will discuss possibilities to test part of the available parameter space at the LHC. 

% -------------------------------------------------------
\subsection{Couplings}
% -------------------------------------------------------

\begin{figure}
    \centering
    \includegraphics[width=0.49\textwidth]{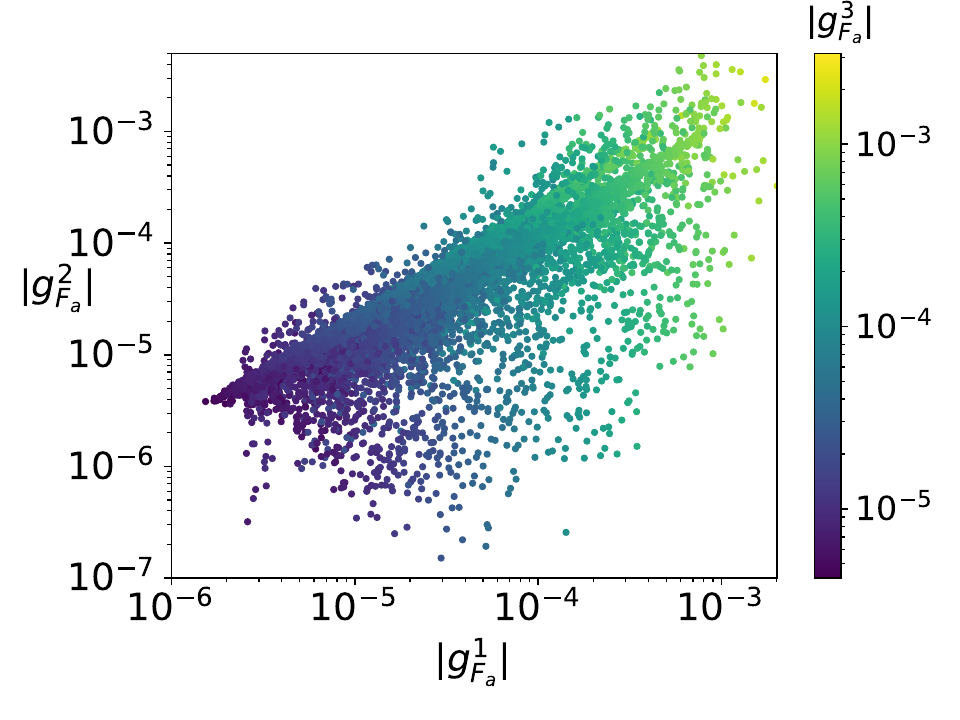}
    \includegraphics[width=0.49\textwidth]{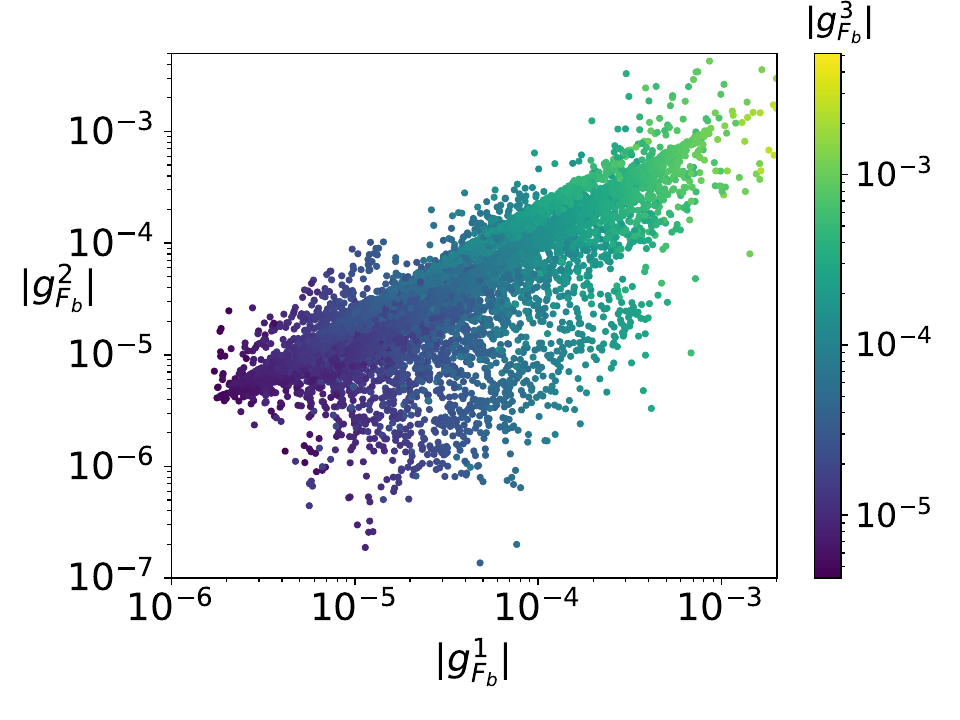}
    \includegraphics[width=0.49\textwidth]{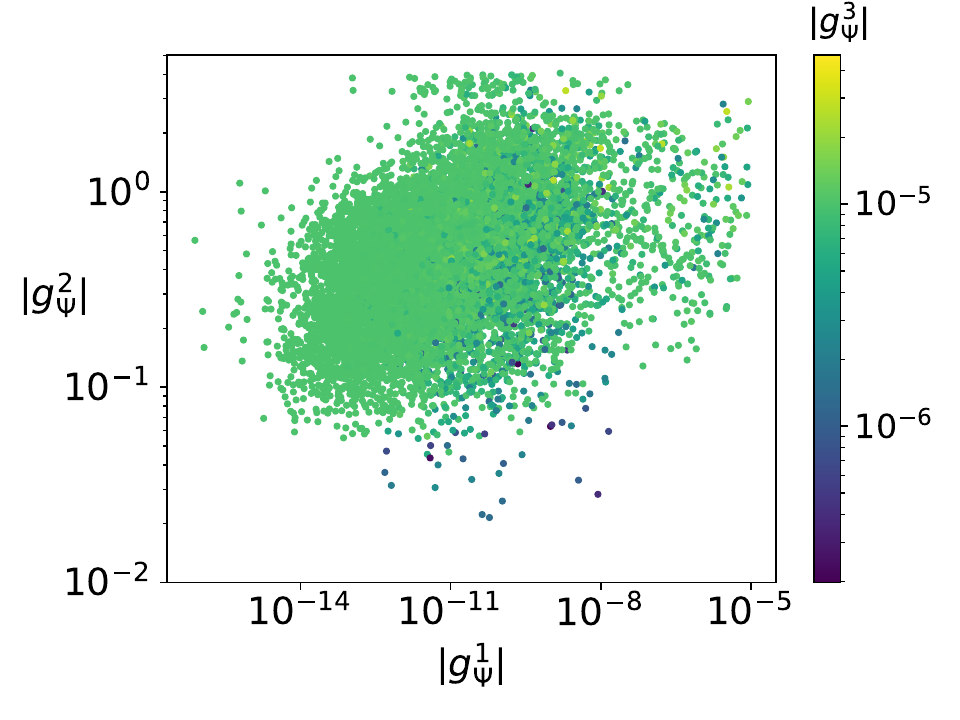}
    \includegraphics[width=0.49\textwidth]{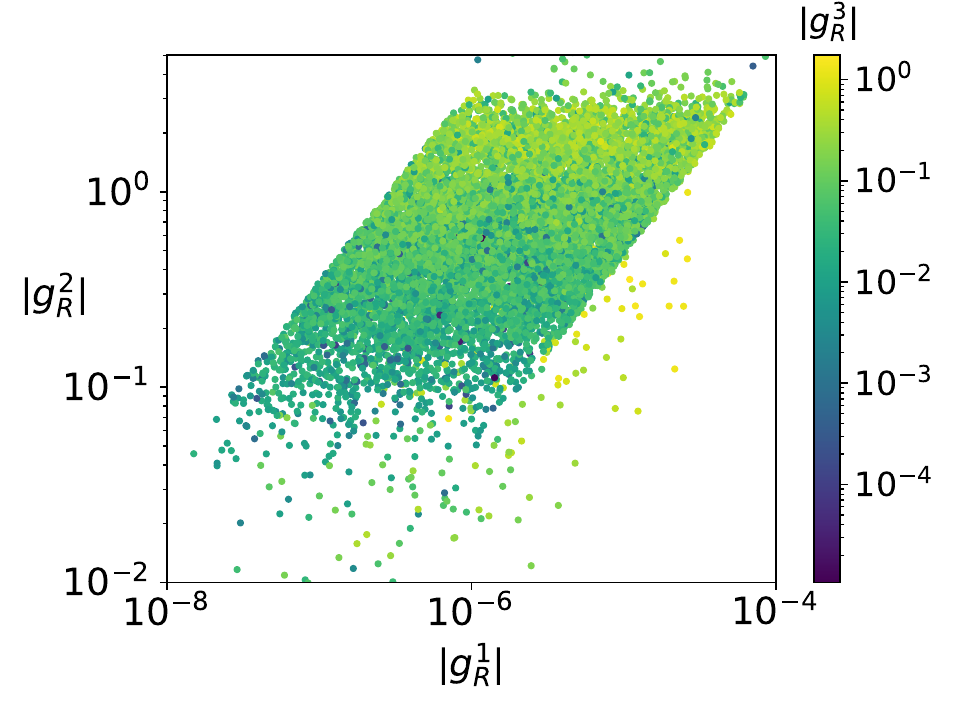}
    \caption{Distributions of the absolute values of the components of the Yukawa couplings $g_{F_1}$ (upper left), $g_{F_2}$ (upper right), $g_{\Psi}$ (lower left) and $g_{R}$ (lower right) obtained from the MCMC scan. The plots show the hierarchy among the components enforced by the neutrino mass fit, the accommodation of $(g-2)_\mu$, and the constraints coming from charged LFV processes.}
    \label{Fig:Couplings1}
\end{figure}

\begin{figure}
    \centering
    \includegraphics[width=0.49\textwidth]{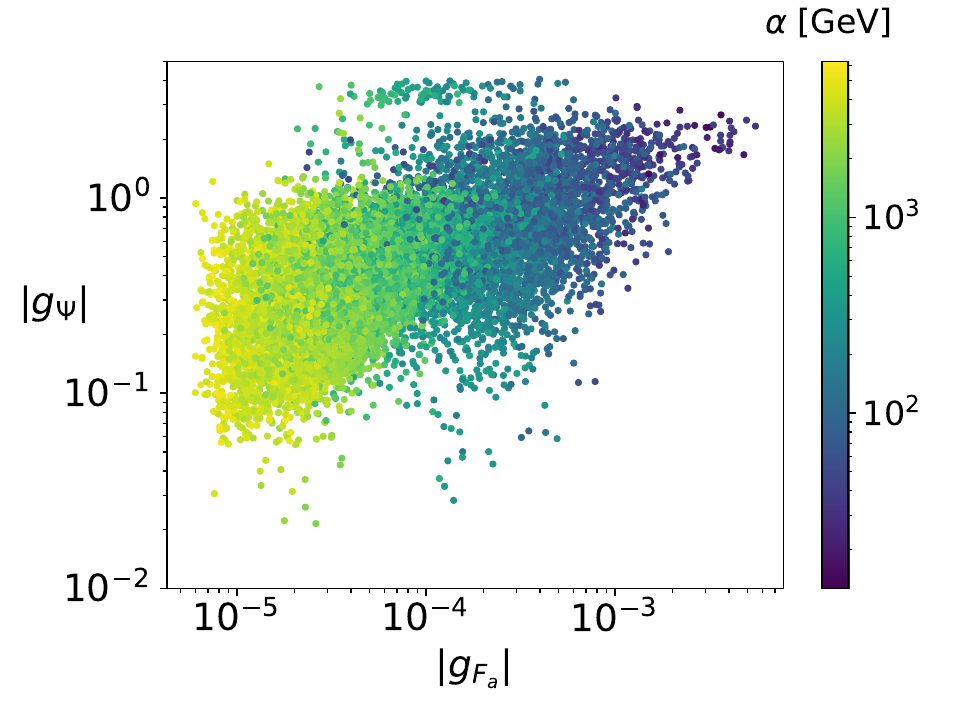}
    \includegraphics[width=0.49\textwidth]{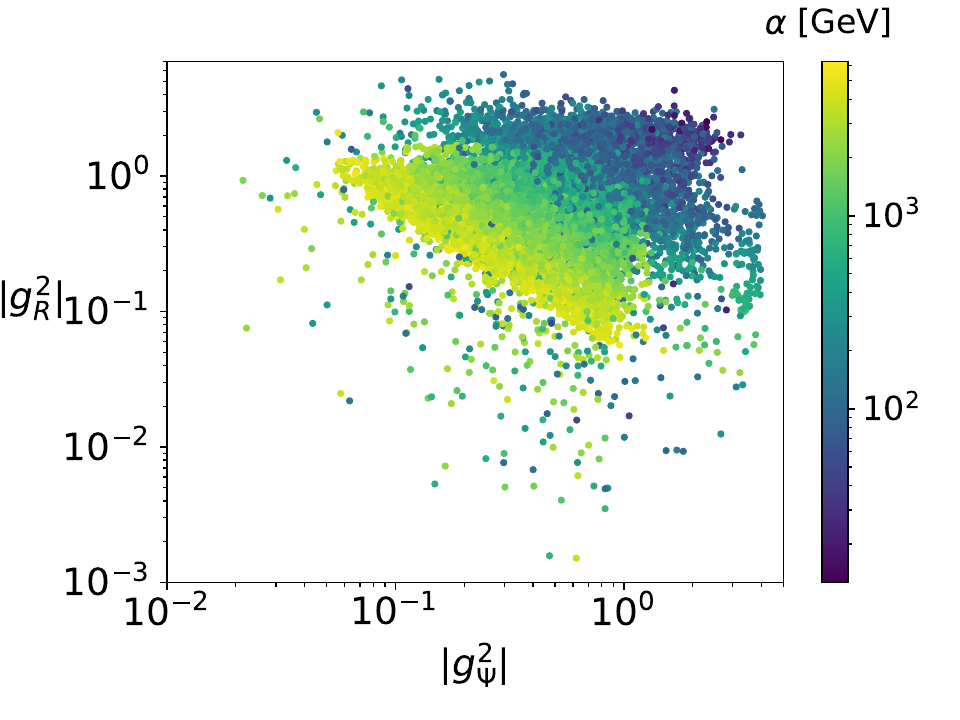}
    \caption{Correlation of selected Yukawa couplings with the trilinear coupling $\alpha$. The couplings $g_\Psi$ and $g_{F_1}$ are connected to the trilinear couplings $\alpha$ through the fit of the neutrino masses, while the connection of $g_\Psi^2$ and $g_R^2$ with $\alpha$ stems from the fit of the anomalous magnetic moment $(g-2)_\mu$.}
    \label{Fig:Couplings2}
\end{figure}

We are interested mainly in the Yukawa couplings that connect the SM particles with the new fields, i.e.\ $g_{F_1}$, $g_{F_2}$, $g_\Psi$, and $g_R$, which are all three-component vectors, see \cref{eq:G_matrix}. These are relevant for neutrino masses, the anomalous magnetic moment and flavour violating decays of the leptons. \cref{Fig:Couplings1} shows the correlations among the different components for each coupling vector, while in \cref{Fig:Couplings2} we show the correlation of selected components with the trilinear coupling $\alpha$.

We clearly see that all components of $g_{F_{1,2}}$ behave in a similar way, with an approximate upper limit of $|g^i_{F_{1,2}}| \lesssim 10^{-3}$ for $i = 1,2,3$. As already explained in \cref{Sec:gminus2}, this upper limit is due to our fit of neutrino masses together with the anomalous magnetic moment $(g-2)_{\mu}$ and the constraints coming from $\mu \to e$ transitions. At the same time, the overall scaling behaviour of all the components of $g_{F_{1,2}}$ is caused by the trilinear coupling $\alpha$, as shown in the left plot of \cref{Fig:Couplings2} for $g_{F_1}$. An analogous behaviour is found for $g_{F_2}$ (not shown). Larger values of $\alpha$ imply a larger scalar mixing, which then suppresses the scale of $g_{F_{1,2}}$ via the neutrino mass fit.

For $g_\Psi$, as already described in \cref{Sec:gminus2} and depicted in \cref{Fig:Gmat_order}, a specific hierarchy among its components is realised to fit the muon anomalous magnetic moment and be below the limits of charged LFV searches. This hierarchy is linked to that of $g_R$, as both contribute equally to these processes, see \cref{Appendix:EMope}. While both $g_\Psi^2$ and $g_R^2$ have to be large to fit $(g-2)_\mu$, $g_\Psi^1$ and $g_R^1$ must remain small to not exceed the current limit of BR($\mu \to e\gamma$). On the same grounds, $g_\Psi^3$ and $g_R^3$ are similarly constrained by the upper limit on BR($\tau \to \mu\gamma$). 

It is worth noting that the fit of $(g-2)_\mu$ links the components of $g_R$ and $g_{\Psi}$ with the trilinear coupling $\alpha$, as can be seen in \cref{Fig:Couplings2} (right). As discussed in \cref{Sec:gminus2}, the dominant contribution to $(g-2)_{\mu}$ and charged LFV decays comes from the left diagram in \cref{Fig:gm2_diagrams}, proportional to $\alpha$. For example, smaller values of $\alpha$ imply larger values of $g_{\Psi}^2$ and $g_R^2$ in order to fit the anomalous magnetic moment $(g-2)_{\mu}$, as can be seen in the upper corner on the right panel of \cref{Fig:Couplings2}. 

The perturbativity requirement for both the Yukawa couplings and $\alpha$ sets then a lower and upper limit on the trilinear coupling of roughly $30~\text{GeV} \lesssim  \alpha \lesssim 4 \, m_{\phi^0_1}$. Note, that the upper bound is actually given for $\alpha/M_\phi$ where $M_\phi$ is the average masses of the scalars involved in this coupling.

% ===========================================================================
\subsection{Charged lepton flavour violating decays}
\label{Sec:clfv}
% ===========================================================================

Charged lepton flavour violating decays rank among the most stringent constraints for neutrino mass models, as fitting the neutrino mixing angles, in general, requires non-diagonal Yukawa matrices that connect also to the charged leptons and allow for transitions between different lepton flavours. While the limits to the branching ratios of these processes are already remarkable, especially for the limit on the decay $\mu \to e\gamma$ from the MEG collaboration \cite{MEG:2016leq}, there is a renovate interest with new experiments expected to take place in the near future, such as MEGII \cite{MEGII:2022}, Mu3e \cite{Blondel:2013ia}, or COMET \cite{COMET:2018auw}, with an expected improvement on the sensitivity of even four orders of magnitude for certain processes like $\mu \to 3e$, or Belle and Belle II for the tau decays \cite{Belle:2010139, Belle:2021ysv, Banerjee:2022vdd}.

\begin{figure}
    \centering
    \includegraphics[width=0.49\textwidth]{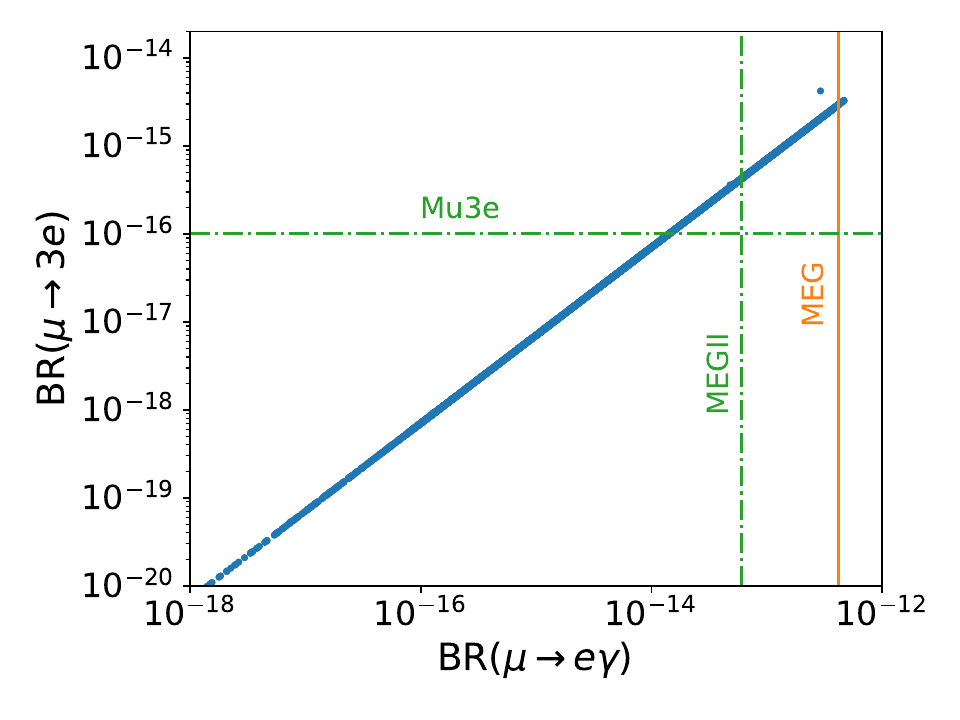}
    \includegraphics[width=0.49\textwidth]{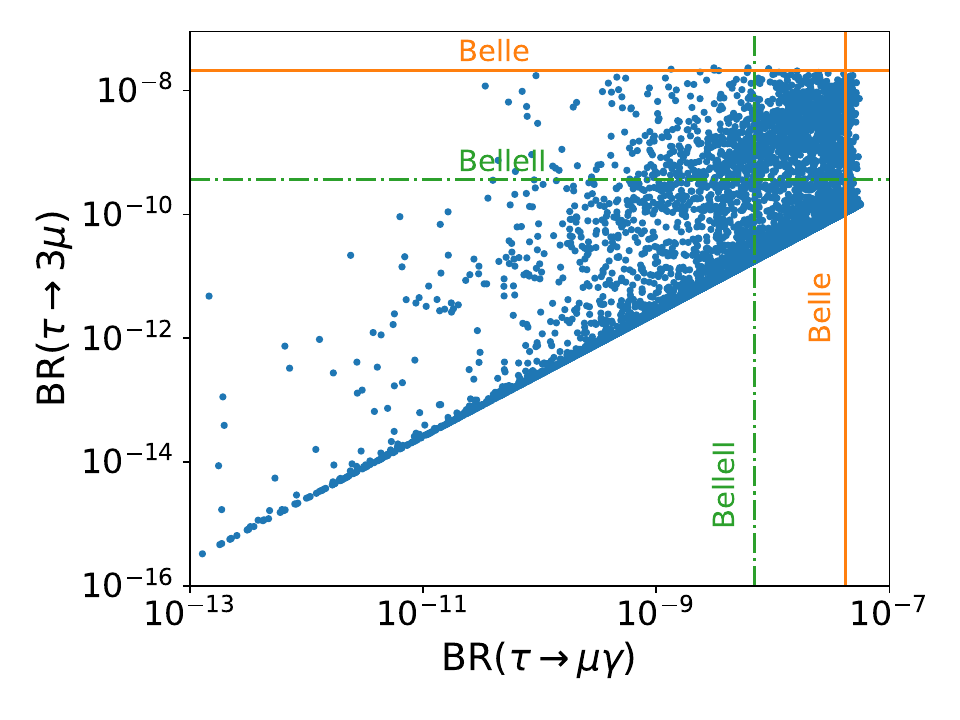}
    \caption{Results for the relevant cLFV decays with the current limits from  MEG collaboration \cite{MEG:2016leq} and Belle \cite{Belle:2010139, Belle:2021ysv} (full lines) and expected sensitivities (dashed lines) from MEGII \cite{MEGII:2022}, Mu3e \cite{Blondel:2013ia} and Belle II \cite{Banerjee:2022vdd}. The other decays not shown here lay below the expected future bounds.}
    \label{fig:clfv}
\end{figure}

Although the charged LFV decays are considered as constraints in our analysis, see \cref{tab:constraints}, it is worth checking how these processes behave in the present model and exploring to which extent future experiments may restrict the parameter space. In \cref{fig:clfv} we show the branching ratios of the most relevant charged LFV decay channels for the muon and the tau, together with their current limits and future expected sensitivities. Note that the muon decays are completely dominated by the dipole contribution, i.e. the diagrams depicted in \cref{Fig:gm2_diagrams}, while there is an important contribution from box diagrams to the tau decays $\tau \to 3 \mu$. This is due to the large values for $g_\Psi^2$ and $g_R^2$ needed to the fit of $(g-2)_\mu$, which makes the box diagram proportional to $g_R^3 \, g_\Psi^{2*} g_R^{2*} g_\Psi^{2}$ dominate over the dipole contribution with the off-shell photon.

For conciseness, we do not show the results for the charged LFV decays of the tau to electrons. For these, we find that, in the best case, the branching ratio of $\tau \to \mu e^+ e^-$ is of the order of $10^{-9}$, just on the border of the future expected sensitivity. For the remaining processes $\tau \to e \gamma$ and $\tau \to 3e$, we obtain branching ratios below $10^{-17}$, not observable in the foreseeable future. The same holds true for the other LFV observables listed in \cref{tab:constraints}.

% ===========================================================================
\subsection{Dark matter observables}
\label{Sec:DM}
% ===========================================================================

Let us recall that the model under consideration includes three possible candidates for cold dark matter (CDM), the lightest $\mathbb{Z}_2$-odd neutral fermion $\chi_1^0$, the lighter scalar $\phi^0_1$, and the pseudo-scalar $A^0$, depending on the mass hierarchies in a given parameter configuration. The dark matter relic density is taken as a constraint in our MCMC analysis, with a theory uncertainty of 10\%, such that $\Omega_{\rm CDM}h^2 = 0.120 \pm 0.012$ \cite{Planck}.

\begin{figure}
    \centering
    \includegraphics[width=0.65\textwidth]{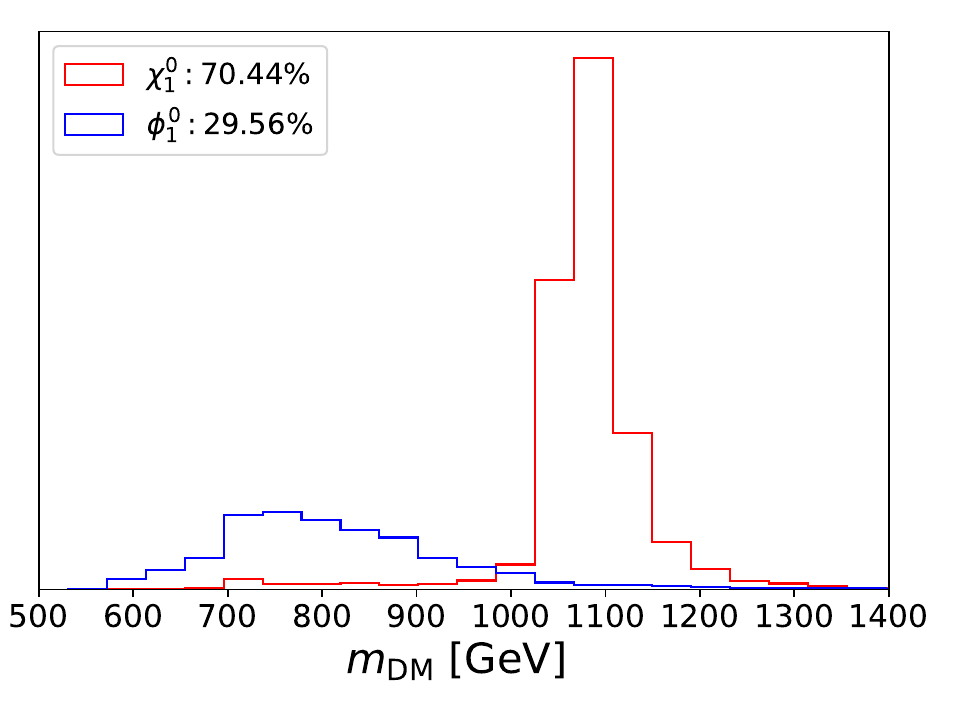}
    \caption{Histograms of the mass and nature of the dark matter candidate. The separation into fermionic and scalar dark matter candidates clearly exhibits a preference for fermionic dark matter with a mass around 1100 GeV.}
    \label{fig:DM_mass}
\end{figure}

Starting with the overall situation, \cref{fig:DM_mass} shows the obtained distribution for the DM mass, separating fermionic ($\chi^0_1$) and scalar ($\phi^0_1$) DM. The shown results exhibit similar behaviour to that found in Ref.\ \cite{Sarazin:2021nwo} for the simpler T1-2-A scotogenic model. Fermionic DM dominates the model parameter space, with a preferred mass of around 1100 GeV. Scalar DM accounts for about 30\% of the viable parameter points, with a preferred masses of roughly 600 GeV to 1000 GeV. 

\begin{figure}
    \centering
    \includegraphics[width=0.49\textwidth]{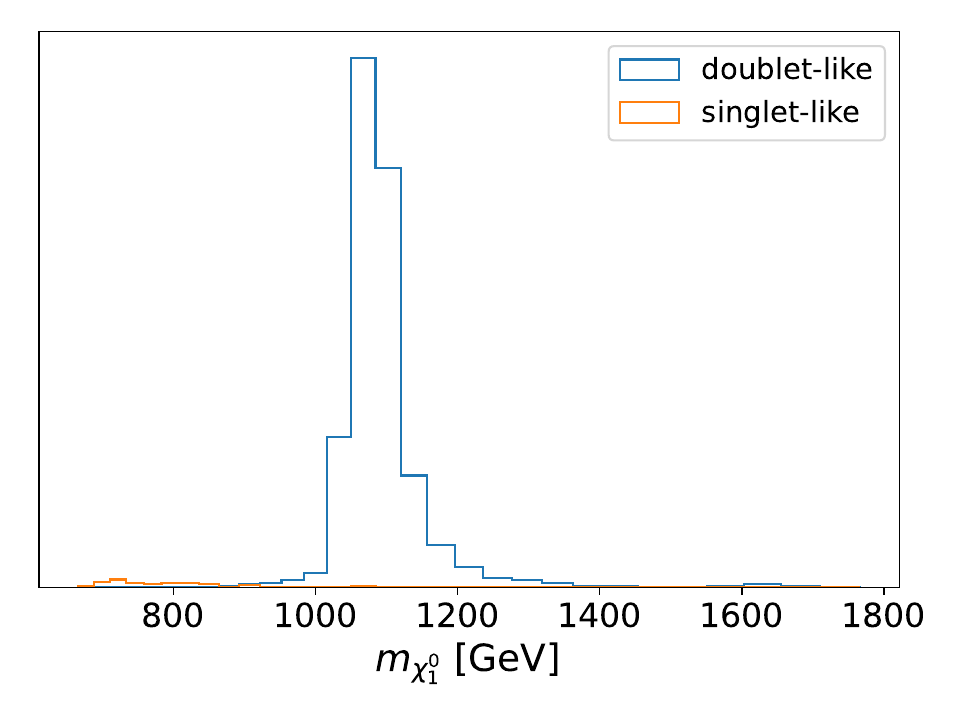}
    \includegraphics[width=0.49\textwidth]{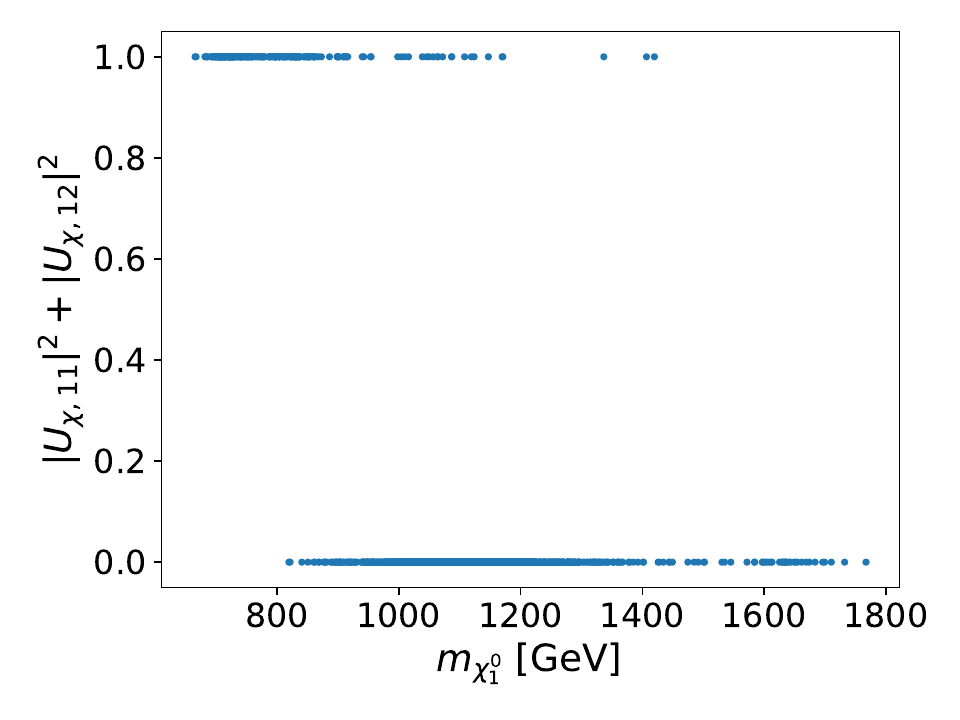}
    \caption{Left: Distribution of the masses in case of fermionic DM candidates, separating the scenarios where the DM candidate is doublet-dominated (blue line) from those where it is singlet-dominated (orange line). Right: singlet content of the DM candidate as a function of the DM mass. }
    \label{fig:DM_fermion}
\end{figure}

As in the T1-2-A model \cite{Sarazin:2021nwo}, fermionic DM is essentially doublet-dominated. This can be seen from \cref{fig:DM_fermion} showing the split into doublet and singlet-dominated DM candidates together with the singlet content of the fermionic DM. This feature can be traced to necessary co-annihilations which occur naturally in the doublet case between $\chi^0_1$ and $\chi^{\pm}$ and $\chi^0_2$ due to the very small mass splitting between these states. The (co-)annihilation processes are dominated by gauge interactions similar to the case of pure higgsinos in supersymmetric models. As already mentioned, sizeable Yukawa couplings to the muons are necessary to explain the potential deviation of its anomalous magnetic moment. This gives additional annihilation channels into muons via scalars, stemming from the doublet $\eta$, in the $t$-channel.

In the case of a fermionic singlet-dominated state $\chi^0_1$, it is much harder to satisfy the relic density requirement from Planck data \cite{Vicente:2014wga}. While singlet fermions $F_i$ are produced thermally, they can only annihilate via the Yukawa $g_{F_i}$ or through the mixing with $\Psi_{1,2}$, both small because of the charged LFV constraints and our fit of $(g-2)_\mu$.

\begin{figure}
    \centering
    \includegraphics[width=0.49\textwidth]{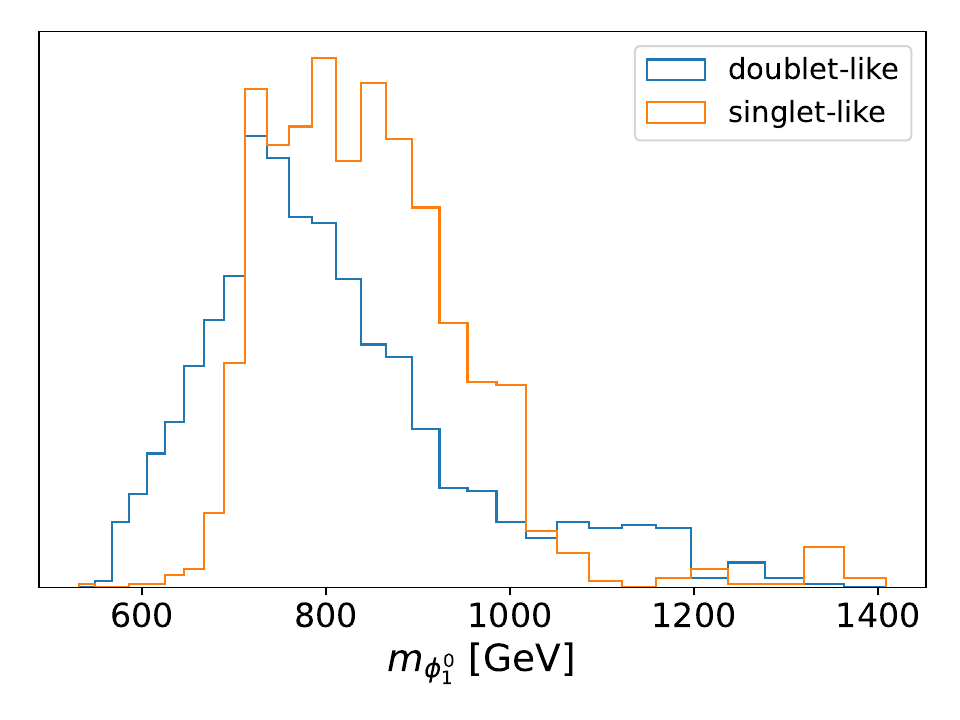}
    \includegraphics[width=0.49\textwidth]{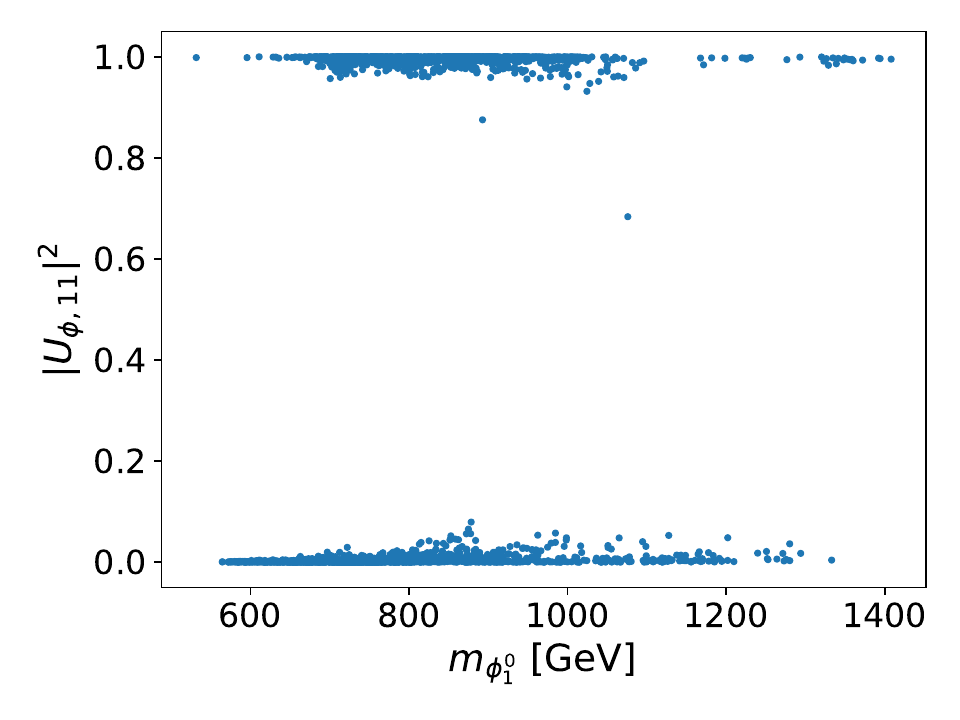}
    \caption{Left: Distribution of the masses in case of scalar DM candidates, separating the scenarios where the DM candidate is doublet-dominated (blue line) from those where it is singlet-dominated (orange line). Right: singlet content of the DM as a function of the DM mass.}
    \label{fig:DM_scalar}
\end{figure}

In the case of scalar DM, there is no clear preference for doublet or singlet-like states within the phenomenologically viable parameter regions, where the preferred masses are typically around 800 GeV. The situation in the scalar sector is more involved compared to fermionic DM. Multiple co-annihilation channels are actively competing, including the Higgs channel, which is significantly constrained by direct-detection limits, as we will discuss subsequently. Additionally, it is worth noting that the associated mass peak exhibits a broader width in the scalar case due to the possibility of larger mass splittings, in contrast to the fermionic case.

\begin{figure}
    \centering
    \includegraphics[width=0.7\textwidth]{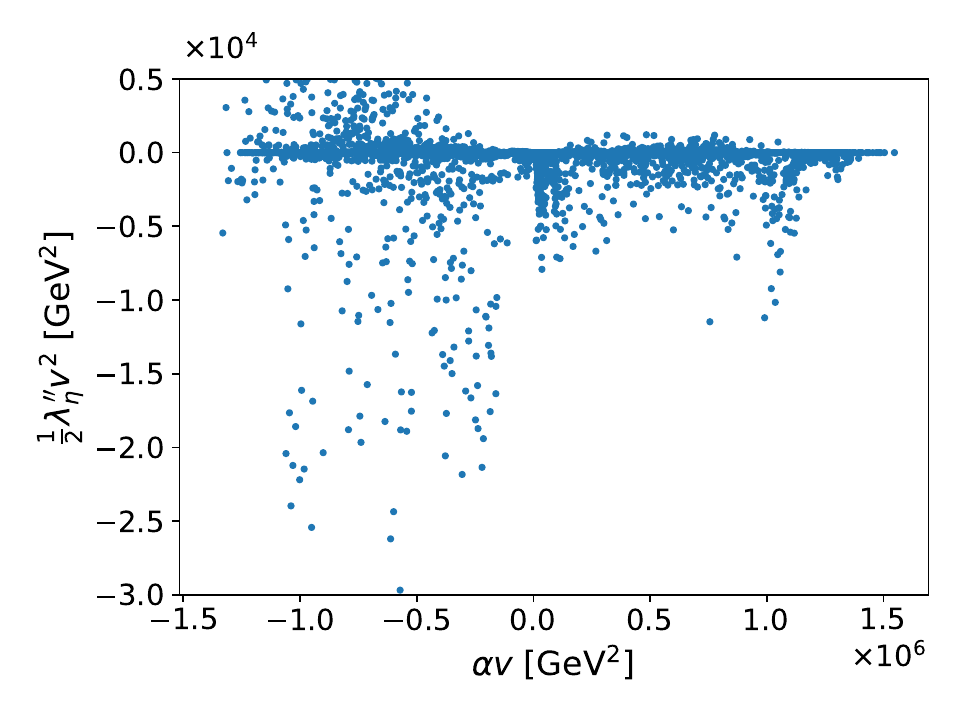}
    \caption{Contributions to the scalar mass matrix in equation \eqref{eq:scalar_mass_matrix}.}
    \label{fig:alphav_vs_Letappv2}
\end{figure}

Finally, we note that we do not find any pseudoscalar DM in this model in contrast to Ref.\ \cite{Sarazin:2021nwo}. The reason can be easily understood by inspecting the mass matrix given in \cref{eq:scalar_mass_matrix}. The mass splitting between the scalar doublet component and the pseudoscalar is given by $\frac{1}{2} \lambda''_{\eta} v^2$ and the mixing between the doublet scalar and the singlet is given by $|\alpha v|$. For a pseudoscalar DM candidate, one needs the doublet to be lighter than the singlet. As can be seen in \cref{fig:alphav_vs_Letappv2} the mixing between the scalar components is always larger than the mass splitting between scalar and pseudoscalar implying that the scalar will be lighter than the pseudoscalar. Note that this feature is less pronounced in the T1-2-A model studied in Ref.\ \cite{Sarazin:2021nwo}, as in their study the trilinear coupling $\alpha$ is rather restricted and, in addition, the constraint from $(g-2)_{\mu}$ has not been taken into account.

\begin{figure}
    \centering
    \includegraphics[width=0.7\textwidth]{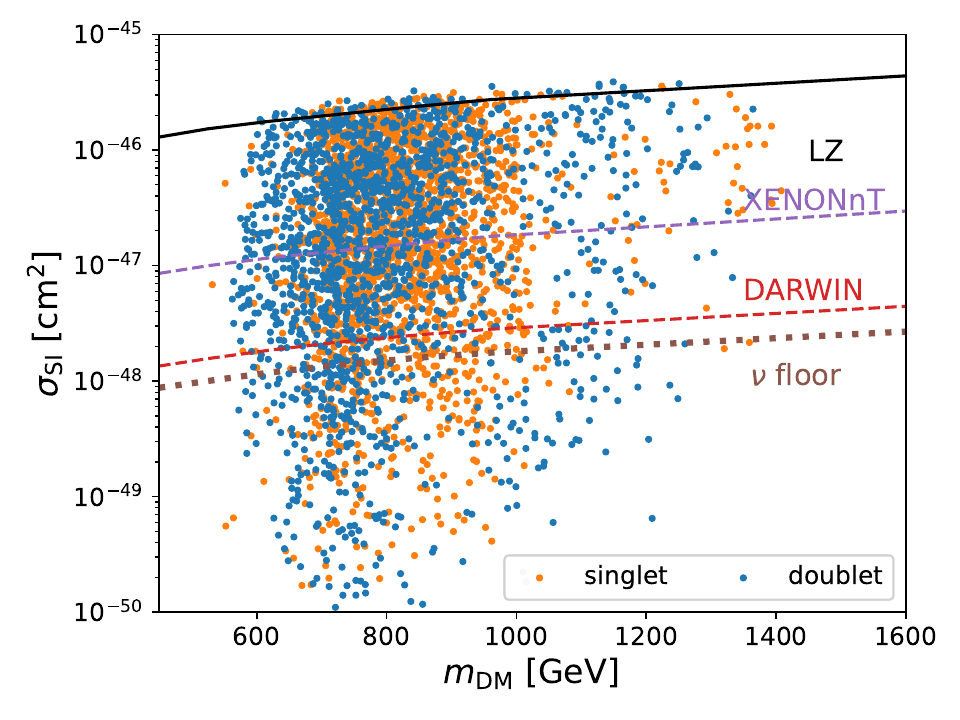}
    \caption{Spin-independent direct detection cross-section versus the mass of the DM in the scalar case, differentiating between singlet (orange) and double-like (blue). The current limit from LUX-ZEPLIN (LZ) \cite{LZ:2022ufs}, as well as the future limits from XENONnT \cite{XENONnT2020} and DARWIN \cite{DARWIN:2016hyl} are given, along with the corresponding line for the neutrino floor \cite{OHare:2021utq}. The fermionic DM case is not shown as the DD cross-section lays below the neutrino floor, around $10^{-60}$ cm$^2$. See text for details.}
    \label{fig:DD_cs}
\end{figure}

In \cref{fig:DD_cs} we show the results for the spin-independent direct detection cross-section for the scalar DM case. As already said in \cref{Sec:observables}, the LZ limit \cite{LZ:2022ufs} is taken as a constraint, such that points not satisfying the current limits are excluded. Most of the remaining viable points can be tested by future experiments like DARWIN \cite{DARWIN:2016hyl}. On the other hand, we do not find any constraints in the case of fermionic dark matter. The reason is that our scan requires the modulus of the relevant Yukawa couplings, $|y_{ij}|$, to be smaller than $10^{-4}$, suppressing the dominant contribution and pushing the cross-sections well below the neutrino floor.

We note that, in both cases, the direct detection cross-section is mainly dominated by Higgs exchange, since we actually have an inelastic dark matter candidate. Inelastic dark matter refers to DM candidates with a mass splitting between the $CP$-even and $CP$-odd components of a neutral state. As the $Z$-boson couples always between the $CP$-even and $CP$-odd components, for the part of the parameter space where the mass splitting between these two states is larger than the kinetic energy of the DM, the contribution from $Z$ channel to the DD cross-section is kinematically forbidden. Since the coupling of the DM to the $Z$-boson has typically gauge strength, if it is active, then it will be excluded by direct detection. We note here that this contribution was added by hand as \verb|micrOMEGAS| does not include inelastic channels. Nevertheless, this excluded very few points, as in practice given the typical DM average relative velocity, the mass splitting needs to be only larger than $\mathcal{O}(100)$ keV to kinematically close the $Z$-channel \cite{Bottaro:2022one}.

% ======================================================
\subsection{Collider aspects}
\label{Sec:Collider}
% ======================================================

\begin{figure}
    \centering
    \includegraphics[width=0.65\textwidth]{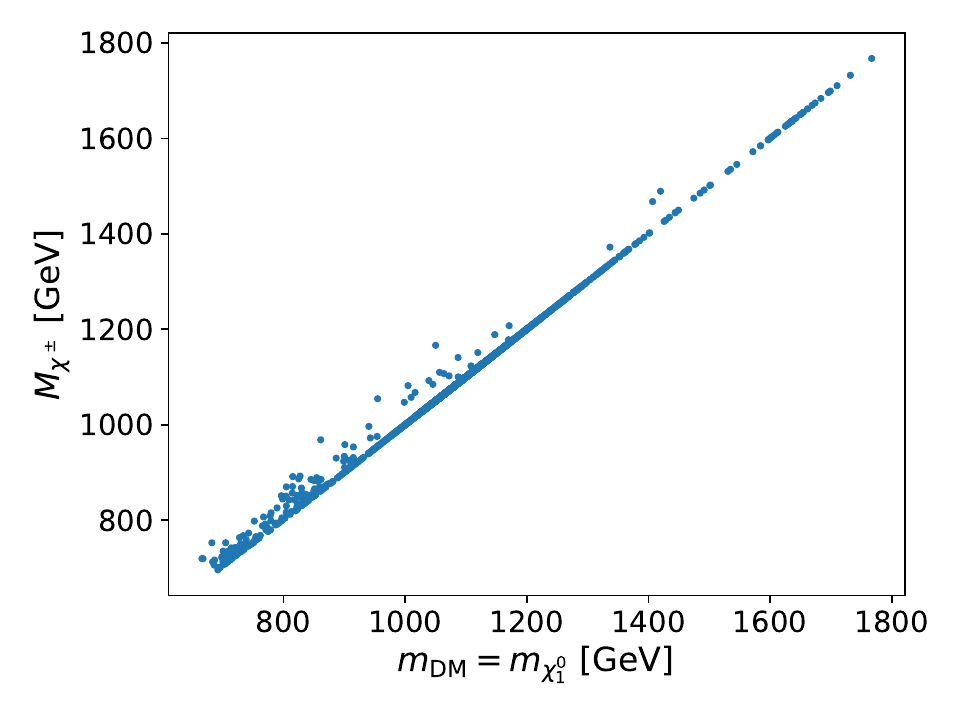}
    \caption{Mass of the charged fermion $\chi^\pm$ as function of the mass of the neutral fermion in the scenarios with a fermionic DM candidate.}
    \label{Fig:DMferm_vs_Charged}
\end{figure}

We have seen in \cref{Sec:DM} that the preferred range for fermionic DM is between 700~GeV and 1.4~TeV with most of the points having a mass between 1 and 1.2~TeV. Moreover, they are essentially always $SU(2)_L$ doublets with the same quantum numbers as higgsinos in supersymmetric models. From Ref.~\cite{LHCxsection} we can thus infer that the cross-section  $\sigma(pp\to \chi^+ \chi^0)$ at the LHC with $\sqrt{s} = 14$~TeV varies between 0.43 fb (1 TeV) and 0.14 fb (1.2 TeV) assuming that both states have the same mass. The signatures depend on the mass difference between the charged and the neutral state which is displayed in \cref{Fig:DMferm_vs_Charged}. In case the DM candidate stems from $SU(2)_L$, one finds only a small difference and the dominant decay mode is $\chi^+ \to \pi^+ \chi^0$. We can infer from the corresponding supersymmetric scenarios that the LHC will not be able to discover the corresponding states, see for example \cite{Barducci:2015ffa} and references therein. In case the DM candidate is singlet-like we find mass splittings between 10 and 150 GeV. The main decay modes are via off-shell neutral scalars
\begin{align}
      \chi^+ \to \ell_R^+ \nu \chi^0_1 \,. 
\end{align}
The interesting point is that the requirement of explaining the $(g-2)$ of the muon implies that nearly all cases one has a muon in the final state. While this is a potentially interesting final state, one should keep in mind that the case of singlet fermionic DM seems to be rarely realised in this model, see \cref{Sec:DM}.

\begin{figure}
    \centering
    \includegraphics[width=0.48\textwidth]{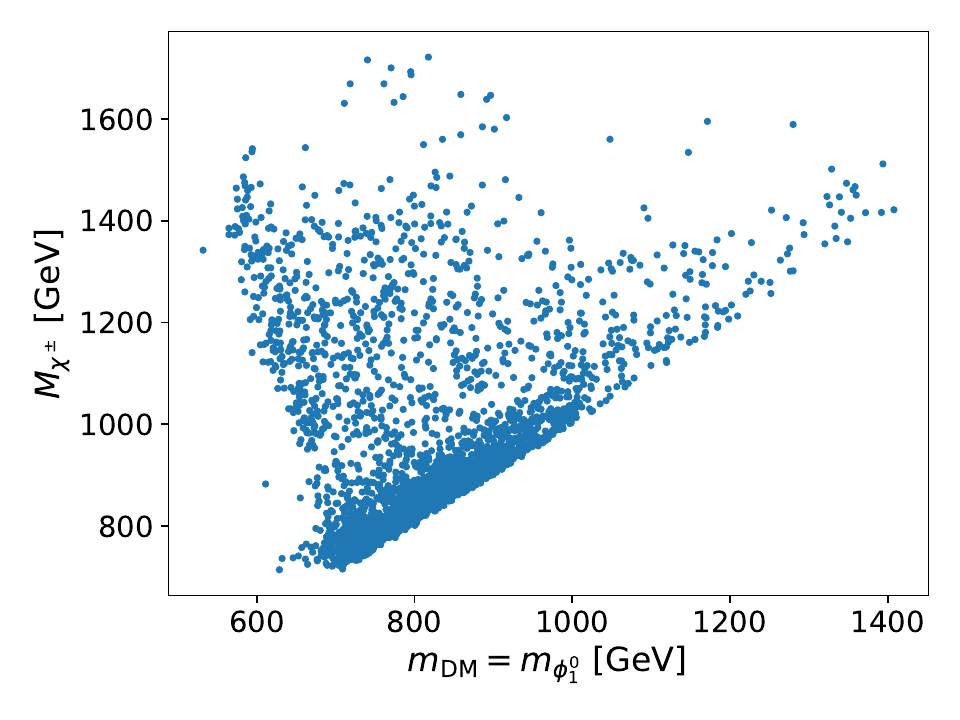}~~~    
    \includegraphics[width=0.48\textwidth]{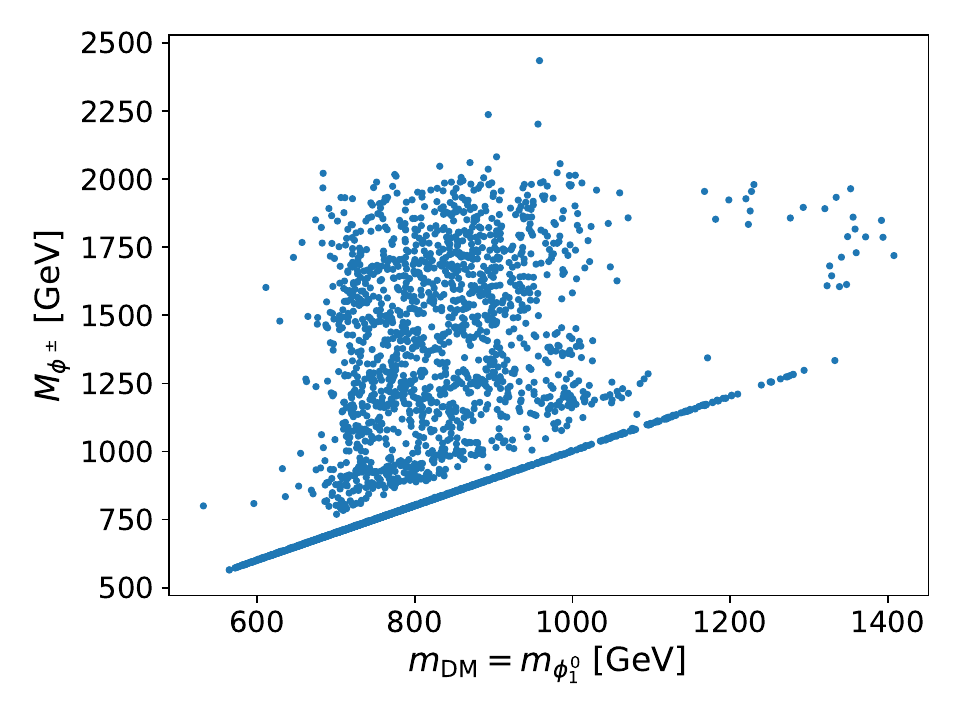}
    \caption{Scalar DM mass versus the mass of the charged fermion $\chi^\pm$ (left) or the mass of the charged scalar $\phi^\pm$ (right).}
    \label{Fig:DMscal_vs_Charged}
\end{figure}

In scenarios with a scalar DM candidate, the situation looks somewhat more promising. We see from the left of \cref{Fig:DMscal_vs_Charged} that in a large portion of the corresponding parameter space the charged fermion has a significantly larger mass than the DM candidate. Note, that the $SU(2)_L$ have a similar mass as the charged fermion. Both will decay into SM-leptons and a $\mathbb{Z}_2$-odd scalar,
\begin{align}
    \chi^+ &\to \nu_L \, \eta^+ \,,\,\, \ell^+ \phi^0_i \\
    \chi^0_j &\to \nu_L \, \phi^0_i \,,\,\, \ell^\pm \eta^\mp
\end{align}
where the $j$ corresponds mainly to the mass eigenstates which are dominantly $SU(2)_L$ fermions. As above, we expect the decays into muons to be dominant. Thus, the signal will dominantly consist of muons in combination with missing transverse energy.

We note for completeness, that one has of course also direct production of scalar doublets. However, already with a mass of about 700 GeV the cross-section is about 0.1~fb  as can be inferred from the production of left-sleptons \cite{LHCxsection} which have the same quantum numbers. We see from the left of \cref{Fig:DMscal_vs_Charged} that $m_{\eta^\pm} \gsim 700$~GeV in scenarios with a sizeable mass splitting in the scalar sector. Thus, the direct production will hardly contribute to an LHC signal for this model.

% ===========================================================================
\section{Leptogenesis}
\label{Sec:Leptogenesis}
% ===========================================================================

This model features heavy Majorana fermions, lepton number violation and complex couplings which are all ingredients for leptogenesis.  In this section, we investigate to which extent one could also explain the observed baryon asymmetry of the Universe via the leptogenesis mechanism\footnote{For a review see, for example, \cite{Buchmuller:2004nz}.}, in the region of parameter space discussed in the previous section. We present here the main results and collect in \cref{App:leptogenesis} further details. 

\begin{figure}
    \centering
    \includegraphics[width=0.32\textwidth]{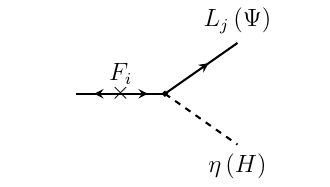} 
    \hfill
    \includegraphics[width=0.32\textwidth]{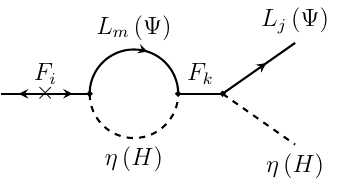}
    \hfill
    \includegraphics[width=0.32\textwidth]{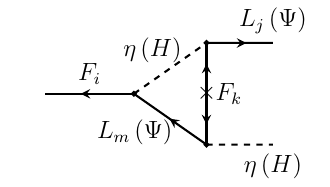} \\[5mm]
    \includegraphics[width=0.32\textwidth]{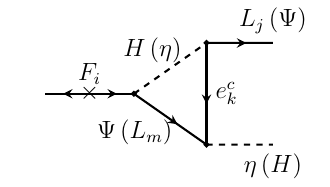}
    \qquad 
    \includegraphics[width=0.32\textwidth]{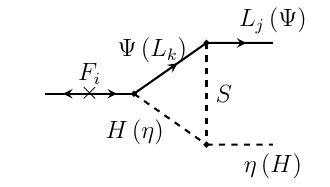}
    \caption{Diagrams contributing to the $CP$ asymmetry generated in the decays of $F_i$ ($i=1,2$). Upper row: diagrams that are similar to the ones obtained in the type-I seesaw model. The arrows indicate the flow of lepton number. Note that another self-energy diagram exists with the mass flip in the internal $F$ and reversing the arrow of $L_m \, (\Psi)$. Lower row: Additional vertex diagrams contributing to the $CP$ asymmetry generated in the decays of the singlet fermions $F_i$. These diagrams in the lower row feature the couplings $g^i_{R}$ (left) and $g^i_{\Psi}$ and the trilinear coupling $\alpha$ (right) which plays an important role in contributing to $(g-2)_{\mu}$, as discussed in Sec.\ \ref{Sec:gminus2}.}
\label{fig:vanilla_lepto}
\end{figure}

We are in a region of parameter space where the couplings $y_{ij}$, which determine the mixing between the $SU(2)_L$ doublet and singlet fermions in \cref{eq:fermion_mass_matrix}, are small. Thus, in practice, only the singlet-like fermions will contribute to a possible lepton asymmetry. These states decay dominantly according to
\begin{align}
 F_i &\to \eta L\,,\,\eta^\dagger \bar{L} \quad \text{ and } \quad 
 F_i \to H \Psi\,,\, H^\dagger  \bar{\Psi}  \,.
\end{align}
The decays will occur at a mass scale which is significantly above the scale of electroweak symmetry breaking and, thus, it is more convenient to work in the gauge basis. At tree-level the former is governed by the couplings $g^i_F$ and the latter by the couplings $y_{ij}$,
\begin{align}
\label{eq:Fdecaywidths}
    &\Gamma(F_i \rightarrow L\,\eta) = \Gamma(F_i \rightarrow \bar{L}\,\eta^{\dag}) ~=~ \displaystyle \sum_{j} \frac{|g^{j}_{F_i}|^2}{32\pi} M_{i} \left(1-\frac{M^2_{\eta}}{M_{i}^2}\right)^2 \\[2mm]
    &\Gamma(F_i \rightarrow \psi\,H) = \Gamma(F_i \rightarrow \bar{\psi}\,H^{\dag}) ~=~ \nonumber\\ &\qquad \frac{M_i}{32\pi}\,\left\{(|y_{1i}|^2+|y_{2i}|^2)\left[1-\left(\frac{M^2_{\Psi}}{M^2_i}\right)^2\right] - 4\,\text{Re}[(y_{1i})^{\ast} y_{2i}]\,\frac{M_{\Psi}}{M_i}\left(1-\frac{M^2_{\Psi}}{M^2_i}\right) \right\} \,,
\end{align}
where we have neglected the masses of the SM leptons and Higgs boson.
The asymmetry is generated at the one-loop level \cite{Fukugita:1986hr} by the diagrams displayed in \cref{fig:vanilla_lepto}. Similar to the type-I seesaw model \cite{Covi:1996wh, Buchmuller:2004nz}, there are the typical wave-function and vertex diagrams depicted in the upper row involving the other singlet fermion in the loop. Moreover, there are additional possible vertex diagrams, which are depicted in the lower row. These involve additional couplings like $g^k_R$ and $\alpha$ which turn out to be important.

The $CP$ asymmetry parameters $\epsilon_{i}$ are given by
\begin{equation}
  \epsilon_{i} = \frac{\Gamma(F_i \rightarrow L\,\eta) + \Gamma(F_i \rightarrow \psi\,H) - \Gamma(F_i \rightarrow \bar{L}\,\eta^{\dag}) - \Gamma(F_i \rightarrow \bar{\psi}\,H^{\dag})}{\Gamma(F_i \rightarrow L\,\eta) + \Gamma(F_i \rightarrow \psi\,H) + \Gamma(F_i \rightarrow \bar{L}\,\eta^{\dag}) + \Gamma(F_i \rightarrow \bar{\psi}\,H^{\dag})} \,.
  \label{eqn:cpasympar}
\end{equation}
Explicit expressions for the various contributions to $\epsilon_i$ in this model may be found in Appendix \ref{App:leptogenesis}.

The MCMC yields that in general ${\rm max}(|g^k_{F_i}|) \gg {\rm max}(|y_{ij}|)$ in most of the parameter space and in the remaining part, they are of equal size. Thus, we will base our estimations in the text below on $g^k_{F_i}$, but we stress that all parameters were properly taken into account in the numerics.

\begin{figure}
    \centering
    \includegraphics[width=0.49\textwidth]{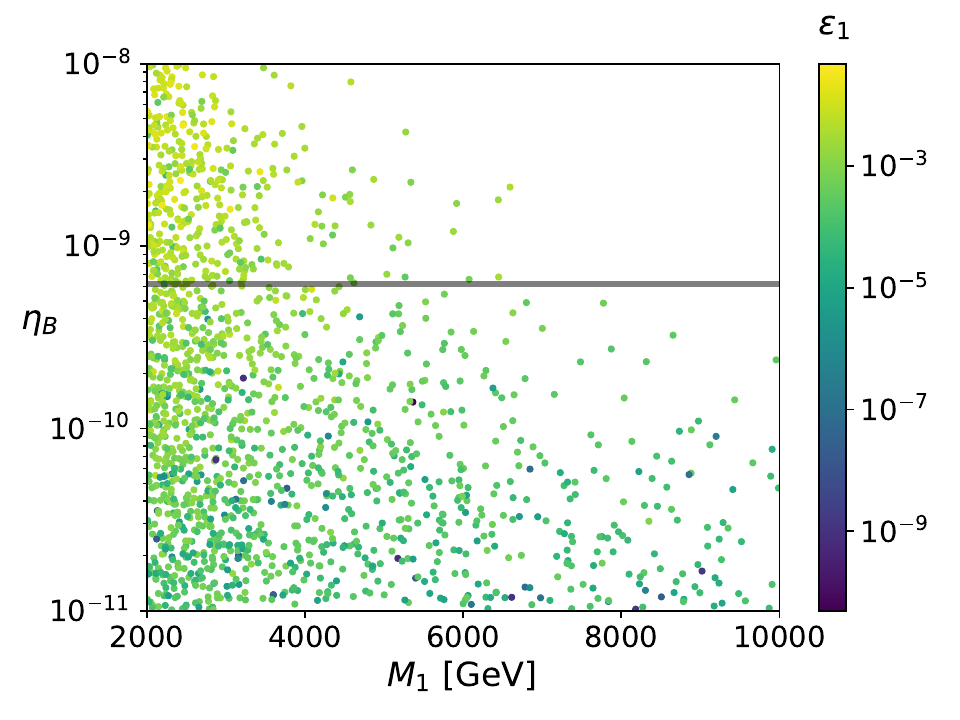}
     \includegraphics[width=0.49\textwidth]{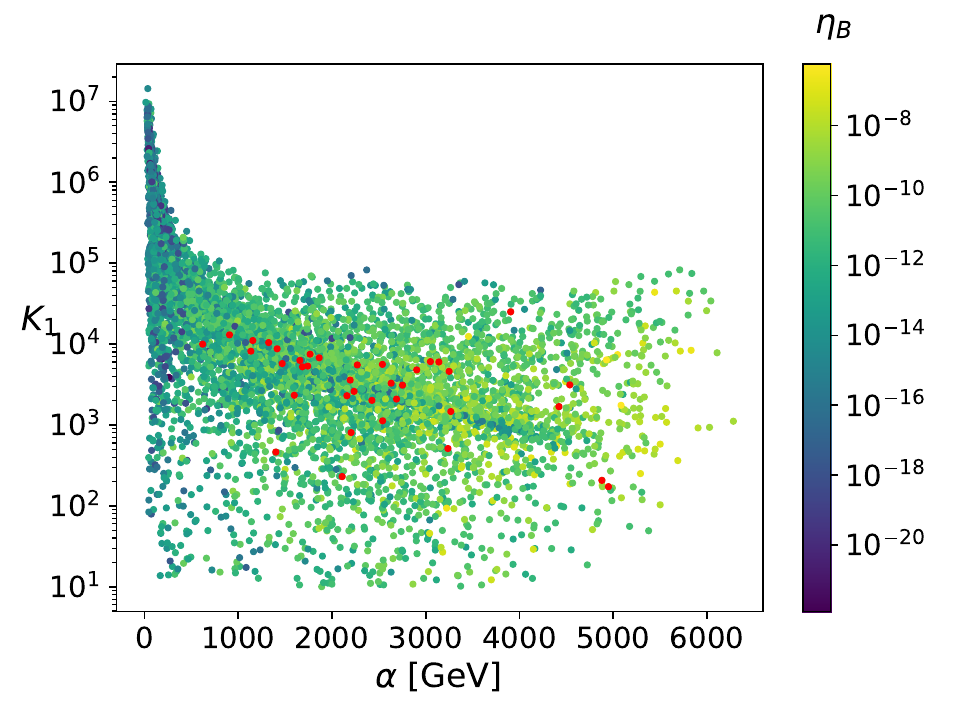} 
    \caption{Left: resulting baryon-to-photon ratio $\eta_B$ plotted against the mass of the lighter singlet fermion driving leptogenesis. The solid grey line denotes the observed value of $\eta_B$ from Planck. The $CP$ asymmetry generated in the decays of the singlet fermion is indicated by the hue. Right: Decay parameter $K_1$ of the lighter singlet fermion versus the absolute value of the trilinear coupling $\alpha$. The value of $\eta_B$ is indicated by the hue. The points in red are within the grey band on the plot to the right.}
    \label{Fig:etaBvsM_CP}
\end{figure}

An important question is to which extent a generated asymmetry gets washed out in the thermal history of the Universe. To answer this, one defines the decay parameters
\begin{equation}
    K_i \equiv \frac{\Gamma_{\text{tot.}}^i}{H(T = M_i)} \simeq  7\times 10^{6} \, \left(\frac{\text{max}\{|g_{F_i}|\}}{10^{-3}}\right)^2 \left(\frac{\text{TeV}}{M_{i}}\right) \,,
    \label{eqn:decaypar}
\end{equation}
with $H$ being the Hubble parameter. The weak washout regime is realised for $K_i \lesssim 1$, the strong washout regime for $K_i \gtrsim 3$, and an intermediate regime in between these values \cite{Buchmuller:2004nz,Davidson:2008bu}. In the parameter space discussed in \cref{Sec:results}, we are always in the strong washout regime, as can be seen from the second part of \cref{eqn:decaypar}. As an example we display $K_1$ in \cref{Fig:etaBvsM_CP}.

The very large values of the decay parameter allow us to neglect washout through scattering processes, as the inverse decays are the dominant sources of the washout. Thus, we may treat leptogenesis as competition between decays and inverse decays \cite{Buchmuller:2004nz}, with the $B-L$ asymmetry being generated when the inverse decays freeze out and the surviving number density $N_{F_i}$ of the $F_i$ decays.  

We solve numerically the corresponding Boltzmann equations given in \cref{App:leptogenesis} with the following initial conditions at high temperatures, 
\begin{equation}
    N_{F_1} ~=~ N^{\text{eq.}}_{F_1}\,,\qquad N_{F_2} ~=~ N^{\text{eq.}}_{F_2}\,, \qquad N_{B-L} = 0 \,,
\end{equation}
and track these number densities down to lower temperatures. As mentioned earlier, we do not assume a large hierarchy between the masses of $F_1$ and $F_2$. Consequently, we take into account both contributions.

Note that the inverse decays of the singlet fermions must freeze-out at a temperature before the sphalerons fall out of equilibrium $(T\sim 100~\text{GeV})$, otherwise, the $B-L$ asymmetry generated in its decay is not converted into baryon asymmetry. As we are in the strong washout regime, we can estimate the freeze-out temperature for the inverse decays as \cite{Giudice:2003jh},
\begin{equation}
    T^{\text{ID}}_{\text{F.O.}} ~\approx~ \frac{M_i}{5 \sqrt{\log(K_i)}}\,.
\end{equation}
This gives an approximate lower bound on the mass of the lightest decaying singlet $M_i \gtrsim 2 \,\text{TeV}$ for which the sphalerons remain active. 
At lower temperatures, $T \ll \text{min}\{M_1,M_2\}$,  we obtain the final $B-L$ asymmetry, $N^{\text{fin.}}_{B-L}$, which is converted to the baryon-to-photon ratio, $\eta_B$, via the sphaleron process \cite{Kuzmin:1985mm,Luty:1992un} using 
\begin{equation}
    \eta_B \equiv \frac{n_B}{n_{\gamma}} ~\approx~ \frac{3}{4}C_{\text{sphal.}} \frac{g^{0}_{\ast}}{g_{\ast}}\,|N^{\text{fin.}}_{B-L}| ~\simeq~ 8 \times 10^{-3} \,|N^{\text{fin.}}_{B-L}|\,.
\end{equation}
$C_{\text{sphal.}}=8/23$ is the sphaleron conversion factor \cite{Harvey:1990qw}, $g^{0}_{\ast} = 43/11$  the present value of the number of relativistic degrees of freedom (DOF) and $g_{\ast}$ the relativistic DOF of the full model at high temperatures.

Figure~\ref{Fig:etaBvsM_CP} depicts the final baryon-to-photon ratio obtained from solving the Boltzmann equations, using the sets of parameters mentioned before, against the mass of the singlet fermion driving leptogenesis. We observe that in contrast to the typical case of strong washout in the type-I seesaw model and the `classic' scotogenic model \cite{Hugle:2018qbw}, the final value of $\eta_B$ has a tendency to decrease with increasing $M_1$. Besides, we also find that the $CP$ asymmetries generated in the decays of lighter $F_1$ are much larger. We note large contributions to $\epsilon_i$ come from the loop diagrams in the lower row of \cref{fig:vanilla_lepto}. 

In the minimal scotogenic model, it is possible to express the decay parameter \eqref{eqn:decaypar} as a function of the lightest neutrino mass and the $\lambda_\eta''$ parameter \cite{Hugle:2018qbw}. In our model, the link is not so direct due to the additional couplings and particle states present. Moreover, the requirement to explain the potential deviation of the $(g-2)$ of the muon while being consistent with the bounds on the LFV lepton decays requires, for example, larger values of the trilinear coupling $\alpha$. As can be seen from the left plot in \cref{Fig:etaBvsM_CP}, the decay parameter decreases with increasing $|\alpha|$. Note that this coupling does not appear directly in the calculation of the decay parameter \eqref{eqn:decaypar}, but reduces the value of the coupling $g_F$ through the neutrino fit, which in turn decreases the tree-level decay width and, hence, lowers the value of $K_i$. 

We also see from the right plot in \cref{Fig:etaBvsM_CP} that only a few points are able to explain the observed baryon asymmetry. Investigating the details of the required parameter combinations to obtain the correct baryon asymmetry would be highly interesting, but is beyond the scope of this paper and left for a future publication. However, one feature that we have observed is that nearly all points present a fermionic doublet as dark matter candidate. Only one out of 25 points in parameter space contains a singlet scalar dark matter candidate. 

% ======================================================
\section{Conclusion}
\label{Sec:Conclusion}
% ======================================================

We have investigated a scotogenic model with a very rich phenomenology. We have presented a complete analysis of the associated parameter space, taking into account constraints from the Higgs sector, the neutrino sector, lepton flavour violating processes, the muon anomalous magnetic moment and dark matter observables. 

Neutrino data governs the couplings of the new particles to the left-handed leptons and the requirement of explaining the observed deviation of the anomalous magnetic moment of the muon $(g-2)$ requires that couplings to muons are sizeable. This in turn implies that  the decays $\mu\to e \gamma$, $\mu\to 3 e$, $\tau\to \mu \gamma$ and $\tau\to 3 \mu$ are in the reach of upcoming experiments in a sizeable part of the parameter space. 

We have found that  the dark matter relic density constraint leads to a preference of fermionic dark matter candidates, in most cases the neutral component of an $SU(2)$ doublet in the mass range 1 to 1.2 TeV. Scenarios featuring a scalar dark matter candidate can be tested by future direct detection experiments like XENONnT or DARWIN, whereas the corresponding cross-sections for fermionic dark matter are well below the so-called neutrino-floor.

We have also briefly discussed the LHC phenomenology in the relevant parameter space. The requirement of explaining the observed deviation in $(g-2)$ leads to a preference for decays into final states containing muons. In particular, in case of a scalar dark matter candidate, we expect muons plus missing transverse energy as dominating signal at the LHC. This signal is also expected in case of supersymmetric models due to the decays of the so-called smuons. However, in our case final states with other leptons or jets in combination with missing transverse energy will be (strongly) suppressed. In case of fermionic dark matter, the mass differences are so small that the charged fermion will decay into a pion and the neutral fermion. The discovery of this final state will be very challenging at the LHC as the required mass imply a relatively low cross-section and, thus, it is likely that the LHC will not be able to observe these particles even in the high luminosity phase.

Finally, we have seen that the available parameter space gets severely constrained if one requires in addition an explanation of the observed baryon asymmetry of the Universe via leptogenesis. We have found that nearly all viable points in the parameter space feature a fermionic dark matter candidate. A detailed analysis of the features of the relevant part in the parameter space will be presented in a future work.

% ===========================================================================
\acknowledgments

The work of M.\,Sarazin is funded by a Ph.D.\ grant of the French Ministry for Education and Research. This work is supported by Campus France/DAAD, project PROCOPE 46704WF, DAAD project number 57561046, and by {\it Investissements d’avenir}, Labex ENIGMASS, contrat ANR-11-LABX-0012. A.\,Alvarez and W.\,Porod are supported by DFG project nr.~PO~1337/8-1. A.\,Banik is supported by DFG, project nr.~HI~744/10-1. R.\,Cepedello is supported by an Alexander von Humboldt Foundation Fellowship. 

% ===========================================================================
\appendix
%============================================================================
\section{New contributions to the electromagnetic dipole moment operator}
\label{Appendix:EMope}
%============================================================================

In this appendix, we collect the additional contributions to the Wilson coefficients of the dipole operator using the notation of Ref.~\cite{Crivellin:2018qmi}. The anomalous magnetic moment, as well as other observables like the electric dipole moment (EDM) and the charged lepton flavour violating (cLFV) decays, is directly connected to the electromagnetic dipole moment operator, i.e.
\begin{equation} \label{eq:cR}
    c_R^{ij} \overline{\ell_j} \sigma_{\mu \nu} P_R \ell_i F^{\mu \nu} + \text{h.c.} \,.
\end{equation}
As already explained in \cref{Sec:gminus2}, the diagonal of the Wilson coefficient $c_R$ is linked to the $(g-2)$ and the EDM by
\begin{equation}
     \overline{\ell_i} \sigma_{\mu \nu}\,\biggr[ \underbrace{\frac{c_R^{ii} + c_R^{ii*}}{2} }_{\text{Re}(c_R^{ii})}+ \underbrace{\gamma^5 \frac{c_R^{ii} - c_R^{ii*}}{2}}_{\gamma^5 \text{Im}(c_R^{ii}) } \biggr]\, \ell_i F^{\mu \nu} \,.
\end{equation}
The real part of the Wilson coefficient will contribute to the anomalous magnetic moment $(g-2)$, whereas the imaginary part is a contribution to the electronic part, the EDM. With this, the anomalous magnetic moment is defined as,
\begin{equation}
    a_i = -4 \frac{m_{\ell_i}}{e} \text{Re}(c_R^{ii}) \,.
\label{Eq:g-2Wilson}
\end{equation}

On the other hand, cLFV decays can be computed directly from \cref{eq:cR}, for which their branching ratio is given by,
\begin{equation}
    \text{BR}(\ell_i \to \ell_j \gamma) = \frac{m_i^3}{4\pi \Gamma_{\ell_i}} \left[ |c_R^{ij}|^2 + |c_R^{ji}|^2 \right] \, ,
\end{equation}
valid for $m_i >> m_j$ and where $\Gamma_{\ell_i}$ is the total decay width of $\ell_i$.

In a general framework, we can consider a coupling between new fermions $\Psi$, new scalars $\Phi$, with the Standard Model lepton $\ell_i$. In this scenario, we have the Lagrangian,
\begin{equation} \label{eq:lagApp}
    \mathcal{L} = \Psi \, \big( \Gamma_L^{i} P_L + \Gamma_R^{i} P_R \big) \, \ell_i \Phi  \, + \, \text{h.c.} \,.
\end{equation}
The associated Wilson coefficients are given by,
\begin{align}\begin{split}
    c_R^{ij} &~=~ \frac{e}{16 \pi^2} \Gamma_L^{i*} \Gamma_R^j M_{\Psi} \frac{f(\frac{M_{\Psi}^2}{M_{\Phi}^2}) + Q g(\frac{M_{\Psi}^2}{M_{\Phi}^2})}{M_{\Phi}^2} \\
    &~~~~~~ + \frac{e}{16 \pi^2} (m_{\ell_j}\Gamma_L^{i*} \Gamma_L^j + m_{\ell_i}\Gamma_R^{i*} \Gamma_R^j) \frac{\Tilde{f}\big(\frac{M_{\Psi}^2}{M_{\Phi}^2}\big) + Q \Tilde{f}\big(\frac{M_{\Psi}^2}{M_{\Phi}^2}\big)}{M_{\Phi}^2} \,
\label{Eq:WilsonCoefficient}   
\end{split}\end{align}
where $Q$ is the electric charge of the fermion in the loop. The functions $f, g, \Tilde{f}, \Tilde{g}$ are defined as,
\begin{align}
    f(x) &~=~ \frac{x^2 - 1 -2x {\log}(x)}{4(x-1)^3} ~=~ 2 \Tilde{g}(x) \,, \\
    g(x) &~=~ \frac{x-1-{\log}(x)}{2(x-1)^2} \,,\\
    \Tilde{f}(x) &~=~ \frac{2x^3 + 3x^2 - 6x +1 -6x^2{\log}(x)}{24(x-1)^4}\,, \\
    \Tilde{g}(x) &~=~  \frac{x^2 - 1 -2x {\log}(x)}{8(x-1)^3} \,.
\end{align}

In order to compute $c_R$, we need to determine the vertices $\Gamma_L^{i}$ and $\Gamma_R^{i}$. In our model, as described in \cref{Sec:Model}, there are two contributions depicted in \cref{Fig:gm2_diagrams}. The Wilson coefficient $c_R$ is then the sum of these two contributions. 

For the first diagram (left) in \cref{Fig:gm2_diagrams}, after EWSB the fields in \cref{eq:lagApp} correspond to $\Psi \equiv \chi^+$ and $\Phi \equiv \phi^0_k$, with couplings
\begin{align}
    \Gamma_{L}^i &~=~ - g_{\Psi}^i (U_{\phi})_{1k} \,, \\
    \Gamma_{R}^i &~=~ \frac{g_R^{i*}}{\sqrt{2}} \left[ (U_{\phi})_{2k} + i \, (U_{\phi})_{3k} \right] \,.
\end{align}
We note here that for simplicity we assumed a sum over the new fields $\Psi$ and $\Phi$ in \cref{Eq:WilsonCoefficient}. Here, a sum over the index $k=1,2,3$ must be performed when computing $c_R$, taking into account that several scalars participate in the diagram, i.e.\ one should replace $M_\Phi$ by $m_{\phi_k^0}$. Also, given our definition of the neutral scalar basis and their mixing as defined in \cref{eq:def_Uphi}, where for conciseness we included the pseudo-scalar $A^0$, here the third $\phi^0$ eigenstate should be considered as $\phi^0_3 \equiv A^0$.

For the second diagram (right) in \cref{Fig:gm2_diagrams}, after EWSB $\Psi \equiv \chi^0_k$ and $\Phi \equiv \eta^-$, with couplings
\begin{align}
    \Gamma_{L}^{i} &~=~  \sum_{j=1}^2 g^i_{F_j} (U_{\chi}^{\dagger})_{jk} \,, \\
    \Gamma_{R}^{i} &~=~ g_R^{i*} (U_{\chi}^{\dagger})_{3k} \,.
\end{align}
Again, a sum over the index $k=1,2,3,4$ must be performed in \cref{Eq:WilsonCoefficient} with $M_\Psi \equiv m_{\chi_k^0}$.

% % ======================================================
\section{Details on the calculation of the baryon asymmetry}
\label{App:leptogenesis}
% % ======================================================

We collect here details of the calculation of the baryon asymmetry of the Universe which has been presented in \cref{Sec:Leptogenesis}.
We largely follow the convention used in \cite{Buchmuller:2004nz}. 

% --------------------------------------------------------
\subsection{Boltzmann equations for leptogenesis}
% --------------------------------------------------------

As a starting point, we write down the Boltzmann equations for thermal leptogenesis, 
\begin{align}
    \frac{dN_{F_i}}{dz_i} &~=~ -K_i\,\frac{z_i\,\mathcal{K}_1(z_i)}{\mathcal{K}_2(z_i)}\big (N_{F_i}-N^{\text{eq.}}_{F_i} \big)\,, 
    \label{eqn:boltzmann_Fdecay}\\
    \frac{dN_{B-L}}{dz_1} &~=~ -\displaystyle \sum_{i=1}^2\,\epsilon_i \,K_i\,\frac{z_i\,\mathcal{K}_1(z_i)}{\mathcal{K}_2(z_i)}(N_{F_i}-N^{\text{eq.}}_{F_i}) - \left(W_{ID}+W_S\right)N_{B-L}\,, 
    \label{eqn:boltzmann_B-L}\\
    &\text{with} \quad  N^{\text{eq.}}_{F_i} =  \frac{z^2_i}{2}\mathcal{K}_2(z_i) \,,\quad W_{ID} = \displaystyle\sum_{i=1}^2 \frac{1}{4}K_i \,z^3_i\, \mathcal{K}_2(z_i)\,,
\end{align}
which track the number densities per co-moving volume $N_{F_i}$ and $N_{B-L}$  of the singlet fermions and the $B-L$ asymmetry. 
Here, we have defined the dimensionless variables $z_i = M_i/T$ where $T$ is the temperature. As the variables are related by $z_2 = z_1 \,(M_2/M_1)$, we solve the above equations in terms of $z_1$. Here we take the contributions of both singlet fermions into account as we do not necessarily have a large mass hierarchy between the two states, see e.g.\ Refs.~\cite{Plumacher:1996kc,Davidson:2008bu}. The quantities $\mathcal{K}_{1,2}(z_i)$ refer to the modified Bessel functions of the second kind and $N^{\text{eq.}}_{F_i}$ refer to the equilibrium distributions of the $F_i$. The term  proportional to $N_{B-L}$ in the second equation is the washout term, and contributions to it are from the inverse decays of the $F_i$, $W_{ID}$, and $\Delta L \neq 0$ scattering processes, $W_S$. The remaining quantities are the decay parameters $K_i$, see \cref{eqn:decaypar}, of the $F_i$ and the $CP$ asymmetry parameters $\epsilon_i$, see \cref{eqn:cpasympar}. The Hubble parameter $H$ enters the calculation of $K_i$. $H$ is given as a function of temperature $T$
\begin{equation}
    H(T) ~=~ \sqrt{\frac{8\pi^3g_{\ast}}{90}}\frac{T^2}{M_{\text{Pl}}} = \frac{H(T = M_i)}{z_i^2}\,,
    \label{eqn:hubblepar}
\end{equation}
where $M_{\text{Pl}} = 1.22 \times 10^{19} \,\text{GeV} $ is the Planck mass. The quantity $g_{\ast}$ refers to the effective number of relativistic degrees of freedom. With the additional particle content, we have $g_{\ast} = 122.25$ compared to the SM value of $g^{\text{SM}}_{\ast} = 106.75$ \cite{Kolb:1990vq}.

We solve the set of coupled Boltzmann equations \eqref{eqn:boltzmann_Fdecay} and \eqref{eqn:boltzmann_B-L} numerically with the following initial conditions at $T = 10^5 M_2$ implying $z_1 \ll 1$ 
\begin{equation}
    N_{F_1} ~=~ N^{\text{eq.}}_{F_1}\,,\qquad N_{F_2} ~=~ N^{\text{eq.}}_{F_2}\,, \qquad N_{B-L} ~=~ 0
    \label{eqn:boltzmann_bc}
\end{equation}

The sphaleron conversion factor enters the calculation of the baryon asymmetry which is given according to Ref.~\cite{Harvey:1990qw} as
\begin{equation}
    C_{\text{sphal.}} ~=~ \frac{24+4 n_D}{66+ 13n_D}
\end{equation}
where $n_D$ is the number of scalar $SU(2)_L$ doublets in the model.

% --------------------------------------------------------
\subsection{Computation of $\epsilon_i$}
% --------------------------------------------------------

Here we collect the expressions for the various diagrams that contribute to the $CP$ asymmetry parameter $\epsilon_i$: 
\begin{equation}
    \epsilon_i ~=~ \sum_j \text{WF}^{(i)}_j + \sum_j \text{V}^{(i)}_j\,.
\end{equation}

Evaluation of the Dirac traces and the required Passarino Veltman reduction of the loop integrals was carried out using FeynCalc \cite{Mertig:1990an, Shtabovenko:2016sxi, Shtabovenko:2020gxv}. The imaginary part of the $B_0$ function is given by
\begin{equation}
    \text{Im}\left[B_0(p^2,m^2_1,m^2_2)\right] ~=~ \pi \frac{\lambda^{\frac{1}{2}}\left(p^2,m^2_1,m^2_2\right)}{p^2} \Theta\left(p^2-(m_1+m_2)^2\right) \, ,
\end{equation}
where $\lambda(x,y,z) = x^2+y^2+z^2-2xy-2yz-2zx$ is the K\"{a}ll\'{e}n function and $\Theta(x)$ is the Heaviside-step function, which enforces the fact that the imaginary part exists only when particles in the loop can go on-shell. The imaginary part of the $C_0$ function was calculated numerically using PackageX \cite{Patel:2016fam}. We indicate all possible sums over the particles in the loop and have summed over the final leptonic states, i.e.\ we do not consider flavour effects. Finally, we have defined 
\begin{equation}
    \Gamma^i_{\text{tot.}} ~=~ \Gamma(F_i \rightarrow L\,\eta) + \Gamma(F_i \rightarrow \psi\,H) + \Gamma(F_i \rightarrow \bar{L}\,\eta^{\dag}) + \Gamma(F_i \rightarrow \bar{\psi}\,H^{\dag}) \,.
\end{equation}

% --------------------------------------------------------
\subsubsection{Wave function diagrams}
% --------------------------------------------------------

\begin{subequations}
\begin{equation}
    \text{WF}_1^{(i)} ~=~ \frac{1}{128\pi^2}\, \frac{M_i}{\Gamma^i_{\text{tot.}}} \,\left(1-\frac{M^2_{\eta}}{M_i^2}\right)^4 \Theta \left(M_i^2 -M^2_{\eta}\right) \displaystyle\sum_{k \neq i}    \text{Im}\left[\left(\displaystyle\sum_{j=1}^3 g^j_{F_i} (g^j_{F_k})^{\ast}\right)^2\right]\frac{M_i M_k}{M^2_i-M^2_k}
\end{equation}
\begin{align}
    &\text{WF}^{(i)}_2 ~=~ \frac{1}{128\pi^2} \frac{M_i\left(1-\frac{M^2_{\eta}}{M^2_i}\right)^2}{\Gamma^i_{\text{tot.}}} \left(1-\frac{M^2_{\Psi}}{M^2_i}\right)\,\Theta\left(M^2_i-M^2_{\Psi}\right)  \nonumber \\
    &\displaystyle\sum_{k\neq  i}\displaystyle\sum_{j=1}^3 \frac{1}{M^2_i-M^2_k}\Bigg\{M^2_{i}\left[\text{Im}\left[g^j_{F_i}\, (g^j_{F_k})^{\ast}\,y_{1k}\,y^{\ast}_{1i}\right]\left(1+\frac{M^2_{\Psi}}{M^2_{i}}\right) + \text{Im}\left[g^j_{F_i}\, (g^j_{F_k})^{\ast}\,y_{1k}\,y^{\ast}_{2i}\right]\frac{M_{\Psi}}{M_{i}}\right]\nonumber  \\[3mm]
    &+ M_i\,M_k\left[\text{Im}\left[g^j_{F_i}\, (g^j_{F_k})^{\ast}\,y_{2k}\,y^{\ast}_{2i}\right]\left(1+\frac{M^2_{\Psi}}{M^2_{i}}\right) + \text{Im}\left[g^j_{F_i}\, (g^j_{F_k})^{\ast}\,y_{2k}\,y^{\ast}_{1i}\right]\frac{M_{\Psi}}{M_{i}}\right] \nonumber \\[3mm]
    &+ M^2_i\left[\text{Im}\left[g^j_{F_i}\, (g^j_{F_k})^{\ast}\,y_{2k}\,y^{\ast}_{2i}\right]\left(1+\frac{M^2_{\Psi}}{M^2_{i}}\right) + \text{Im}\left[g^j_{F_i}\, (g^j_{F_k})^{\ast}\,y_{1k}\,y^{\ast}_{2i}\right]\frac{M_{\Psi}}{M_{i}}\right] \nonumber  \\[3mm]
    &+ M_i\,M_k\left[\text{Im}\left[g^j_{F_i}\, (g^j_{F_k})^{\ast}\,y_{1k}\,y^{\ast}_{1i}\right]\left(1+\frac{M^2_{\Psi}}{M^2_{i}}\right) + \text{Im}\left[g^j_{F_i}\, (g^j_{F_k})^{\ast}\,y_{2k}\,y^{\ast}_{1i}\right]\frac{M_{\Psi}}{M_{i}}\right]\Bigg\}
 \end{align}
\begin{align}
    &\text{WF}^{(i)}_3 ~=~ \frac{1}{128\pi^2}\,\frac{M_i}{\Gamma^i_{\text{tot.}}}\,\left(1-\frac{M^2_{\Psi}}{M^2_i}\right)^2\,\Theta\left(M^2_i-M^2_{\Psi}\right) \nonumber \\[3mm]
    & \quad \displaystyle\sum_{k\neq  i}\frac{1}{M^2_i-M^2_k} \Bigg\{M^2_i\left[\text{Im}\left[y_{1i}\,y_{2i}\,y^{\ast}_{1k}\,y^{\ast}_{2k}\right]\left(1+\frac{M^2_{\Psi}}{M^2_{i}}\right)-2\,\text{Im}\left[(y^2_{1i}+y^2_{2i})\,y^{\ast}_{1k}\,y^{\ast}_{2k}\right]\frac{M_{\Psi}}{M_i}\right] \nonumber \\[3mm]
    &\quad   +M_{\Psi}\,M_i\left[\text{Im}\left[(y^2_{1i}+y^2_{2i})\,y^{\ast}_{1k}\,y^{\ast}_{2k}\right]\left(1+\frac{M^2_{\Psi}}{M^2_{i}}\right)-2\,\text{Im}\left[y_{1i}\,y_{2i}\,y^{\ast}_{1k}\,y^{\ast}_{2k}\right]\frac{M_{\Psi}}{M_i}\right] \nonumber \\[3mm]
    &\quad  + M_k\,M_i\left[\text{Im}\left[y^{2}_{1i}\,(y^{\ast}_{1k})^2+y^{2}_{1i}\,(y^{\ast}_{2k})^2\right]\left(1+\frac{M^2_{\Psi}}{M^2_{i}}\right)-2\,\text{Im}\left[[(y^{\ast}_{1k})^2+(y^{\ast}_{2k})^2]\,y^i_1\,y^i_2\right]\frac{M_{\Psi}}{M_i}\right] \nonumber \\[3mm]
    &\quad  +M_k\,M_{\Psi}\left[\text{Im}\left[[(y^{\ast}_{1k})^2+(y^{\ast}_{2k})^2]\,y_{1i}\,y_{2i}\right]\left(1+\frac{M^2_{\Psi}}{M^2_{i}}\right)-2\,\text{Im}\left[y^{2}_{1i}\,(y^{\ast}_{1k})^2+y^{2}_{1i}\,(y^{\ast}_{2k})^2\right]\frac{M_{\Psi}}{M_i}\right]\Bigg\}
\end{align}
\begin{align}
    &\text{WF}^{(i)}_4 ~=~ \frac{1}{128\pi^2}\frac{M_i\left(1-\frac{M^2_{\Psi}}{M^2_i}\right)}{\Gamma^i_{\text{tot.}}} \left(1-\frac{M^2_{\eta}}{M^2_i}\right)^2\,\Theta\left(M^2_i-M^2_{\eta}\right)  \nonumber \\[3mm]
     & \qquad \displaystyle\sum_{k\neq  i}\displaystyle\sum_{j =  1}^3\frac{1}{M^2_i-M^2_k} \Bigg\{M^2_i\left[\text{Im}\left[g^j_{F_k}\,(g^j_{F_i})^{\ast}\,y_{1i}\, y^{\ast}_{1k}\right]\left(1+\frac{M^2_{\Psi}}{M^2_{i}}\right)-2\,\text{Im}\left[g^j_{F_k}\,(g^j_{F_i})^{\ast}\,y_{2i} \,y^{\ast}_{1k}\right]\frac{M_{\Psi}}{M_i}\right] \nonumber \\[3mm]
    & \qquad + M_i\,M_k\left[\text{Im}\left[g^j_{F_k}\,(g^j_{F_i})^{\ast}\,y_{2i} \,y^{\ast}_{2k}\right]\left(1+\frac{M^2_{\Psi}}{M^2_{i}}\right)-2\,\text{Im}\left[g^j_{F_k}\,(g^j_{F_i})^{\ast}\,y_{1i} \,y^{\ast}_{2k}\right]\frac{M_{\Psi}}{M_i}\right]\nonumber \\[3mm]
    &\qquad +M^2_i\left[\text{Im}\left[g^j_{F_i}\,(g^j_{F_k})^{\ast}\,y_{1i} \,y^{\ast}_{1k}\right]\left(1+\frac{M^2_{\Psi}}{M^2_{i}}\right)-2\,\text{Im}\left[g^j_{F_i}\,(g^j_{F_k})^{\ast}\,y_{2i}\, y^{\ast}_{1k}\right]\frac{M_{\Psi}}{M_i}\right]\nonumber \\[3mm]
    &\qquad + M_i\,M_k\left[\text{Im}\left[g^j_{F_i}\,(g^j_{F_k})^{\ast}\,y_{2i} \,y^{\ast}_{2k}\right]\left(1+\frac{M^2_{\Psi}}{M^2_{i}}\right)-2\,\text{Im}\left[g^j_{F_i}\,(g^j_{F_k})^{\ast}\,y_{1i} \,y^{\ast}_{2k}\right]\frac{M_{\Psi}}{M_i}\right]\Bigg\}
\end{align}
\end{subequations}

% --------------------------------------------------------
\subsubsection{Vertex diagrams}
% --------------------------------------------------------

\begin{align}
    \text{V}^{(i)}_1 &~=~ \frac{1}{128\pi^2} \frac{M_i\left(1-\frac{M^2_{\eta}}{M_i^2}\right)^2}{\Gamma^i_{\text{tot.}}}\nonumber \\ &\displaystyle\sum_{k \neq i} \text{Im}\left[\left(\displaystyle\sum_{j=1}^3 g^j_{F_i} (g^j_{F_k})^{\ast}\right)^2\right] M_k\,M_i\, \Bigg\{\frac{\Theta\left(M^2_i-M^2_{\eta}\right)}{M^2_i}  - \frac{M^2_{\eta} - M^2_k}{M^2_i-M^2_{\eta}}\,\frac{\Theta\left(M^2_{\eta}-M^2_k\right)}{M^2_{\eta}} \nonumber \\[3mm] 
    &\qquad \qquad \qquad  +2\left(1-\frac{M^2_{\eta}-M^2_k}{M^2_i-M^2_{\eta}}\right)\text{Im}\left[\frac{C_0(0,M^2_{\eta},M^2_i,M^2_{\eta},M^2_k,0)}{\pi }\right]\Bigg\}
\end{align}

\begin{align}
    \text{V}^{(i)}_2 &~=~ \frac{1}{128\pi^2}\frac{M_i\left(1-\frac{M^2_{\Psi}}{M^2_i}\right)}{\Gamma^i_{\text{tot.}}} \displaystyle\sum_{k \neq i}\frac{1}{(M^2_i-M^2_{\Psi})^2}\Bigg\{\left(1+\frac{M^2_{\Psi}}{M^2_i}\right)\left(\mathcal{F}_{1a}+\mathcal{F}_{1b}\right)+ 2\frac{M_{\Psi}}{M_i}\left(\mathcal{F}_{2a}+\mathcal{F}_{2b}\right)\Bigg\}
\end{align}
where 
\begin{align*}
    &\mathcal{F}_{1a} ~=~ \\
    &-\left(1-\frac{M^2_{\Psi}}{M^2_i}\right)\Theta\left(M^2_i - M^2_{\Psi}\right)\Bigg\{-2 M_k^2 M_i M_{\Psi}\,\text{Im}\left[ y_{1i}\, y_{2i}\, \left(y_{2k}^*\right)^2\right]-M_k M_i \left(M_i^2+M_{\Psi}^2\right) \text{Im}\left[y_{1i}^2\, \left(y_{1k}^*\right)^2\right]\\[3mm]
    &-M_i^4\, \text{Im}\left[\left(y_{1i}\, y_{2i}\, y_{1k}^* \,y_{2k}^*\right)\right]-M_i^3 M_{\Psi} \,\text{Im}\left[y^2_{1i} \,y_{1k}^*\, y_{2k}^*\right]+3 M_i^2 M_{\Psi}^2 \,\text{Im}\left[y_{1i}\, y_{2i}\, y_{1k}^*\, y_{2k}^*\right]\\[3mm]
    &+2 M_i M_{\Psi}^3 \,\text{Im}\left[y^2_{1i}\, y_{1k}^*\, y_{2k}^*\right]\Bigg\} \\[3mm]
    &+\left(1-\frac{M^2_{k}}{M^2_{\Psi}}\right)\Theta\left( M^2_{\Psi}-M^2_k\right)\Bigg\{-M_k \,M_{\Psi} \left(M_i^2+M_{\Psi}^2\right) \text{Im}\left[y_{1i}\, y_{2i} \,y_{2k}^*\, y_{2k}^*\right]-2 M_k M_i M_{\Psi}^2\, \text{Im}\left[y_{1i}^2 \,\left(y_{1k}^*\right)^2\right] \\[3mm]
    &-M_i^3 \,M_{\Psi} \,\text{Im}\left[y^2_{1i}\,y_{1k}^* y_{2k}^*\right]+3 M_i M_{\Psi}^3 \,\text{Im}\left[y^2_{1i}\, y_{1k}^*\, y_{2k}^*\right]+2 M_{\Psi}^4 \,\text{Im}\left[y_{1i}\, y_{2i}\, y_{1k}^*\, y_{2k}^*\right]\Bigg\} \\[3mm]
    &+\text{Im}\left[\frac{C_0(M^2_{\Psi},0,M^2_i,0,M^2_k,M^2_{\Psi})}{\pi }\right]\Bigg\{-M_k \,M_{\Psi} \left(-2 \left(M_k^2-M_{\Psi}^2\right)-M_i^4+M_{\Psi}^4\right) \text{Im}\left[y_{1i}\, y_{2i}\, \left(y_{2k}^*\right)^2\right] \\[3mm]
    &+M_k^2 M_i^4 \,\text{Im}\left[y_{1i} \,y_{2i}\, y_{1k}^* \,y_{2k}^*\right]-M_k^2 M_i^3 M_{\Psi}\, \text{Im}\left[y^2_{1i}\, y_{1k}^*\, y_{2k}^*\right]+M_k^2 M_i^3 M_{\Psi}\, \text{Im}\left[y^2_{1i}\, y_{1k}^*\, y_{2k}^*\right] \\[3mm]
    &-2 M_k^2 M_i^2 M_{\Psi}^2\, \text{Im}\left[y_{1i} \,y_{2i}\, y_{1k}^*\, y_{2k}^*\right]+M_k^2 M_i^2 M_{\Psi}^2 \,\text{Im}\left[y_{1i}\, y_{2i}\, y_{1k}^*\, y_{2k}^*\right] \\[3mm]
    &-M_k\, M_i \left(M_i^2+M_{\Psi}^2\right) \left(-M_k^2-M_i^2+2 M_{\Psi}^2\right) \text{Im}\left[y^2_{1i}\, \left(y_{1k}^*\right)^2\right]+2 M_{\Psi}^2 \left(M_{\Psi}^4-M_k^2 M_i^2\right) \text{Im}\left[y_{1i}\, y_{2i}\, y_{1k}^*\, y_{2k}^*\right] \\[3mm]
    &-2 M_k^2 M_i\, M_{\Psi}^3\,\text{Im}\left[y^2_{1i}\, y_{1k}^* \,y_{2k}^*\right] + M_k \,M_{\Psi} \left(M_i^2-M_{\Psi}^2\right)^2 \text{Im}\left[y_{1i}\, y_{2i}\, \left(y_{1k}^*\right)^2\right]-M_i^4 M_{\Psi}^2 \,\text{Im}\left[y_{1i}\, y_{2i}\, y_{1k}^*\, y_{2k}^*\right]\\[3mm]
    &+M_i^4 M_{\Psi}^2 \text{Im}\left[y_{1i}\, y_{2i}\, y_{1k}^*\, y_{2k}^*\right]-2 M_i^3 M_{\Psi}^3\text{Im}\left[y^2_{1i}\, y_{1k}^*\, y_{2k}^*\right]-M_i^2 M_{\Psi}^4 \text{Im}\left[y_{1i}\, y_{2i}\, y_{1k}^* \,y_{2k}^*\right] \\[3mm]
    &+4 M_i\, M_{\Psi}^5 \text{Im}\left[y^2_{1i}\, y_{1k}^*\, y_{2k}^*\right]+M_{\Psi}^6 \,\text{Im}\left[y_{1i} \,y_{2i}\, y_{1k}^*\, y_{2k}^*\right]\Bigg\}
\end{align*}

\begin{align*}
    &\mathcal{F}_{1b} ~=~  \\
    &\left(1-\frac{M^2_{\Psi}}{M^2_i}\right)\Theta\left(M^2_i - M^2_{\Psi}\right)\Bigg\{M_k\, M_i\left(M_i^2+M_{\Psi}^2\right) \text{Im}\left[y^2_{2i}\, \left(y_{2k}^*\right)^2\right]+2 M_k\, M_i^2 M_{\Psi} \,\text{Im}\left[\left(y_{1k}^*\right)^2 y_{1i}\, y_{2i}\right] \\[3mm]
    &+M_i^4 \text{Im}\left[y_{1i}\, y_{2i}\, y_{1k}^*\, y_{2k}^*\right]+ M_i^3 M_{\Psi} \,\text{Im}\left[y^2_{2i}\, y_{1k}^*\, y_{2k}^*\right]-3 M_i^2 M_{\Psi}^2 \,\text{Im}\left[y_{1i}\, y_{2i}\, y_{1k}^* \,y_{2k}^*\right] \\[3mm]
    &-2 M_i\,M_{\Psi}^3 \text{Im}\left[y^2_{2i}\,  y_{1k}^*\, y_{2k}^*\right]\Bigg\} \\[3mm]
    &-\left(1-\frac{M^2_{k}}{M^2_{\Psi}}\right)\Theta\left( M^2_{\Psi}-M^2_k\right)\Bigg\{M_k \,M_{\Psi} \left(M_i^2+M_{\Psi}^2\right) \text{Im}\left[\left(y_{1k}^*\right)^2 y_{1i}\, y_{2i}\right]+2 M_k\, M_i\,M_{\Psi}^2\,\text{Im}\left[y^2_{2i}\, y_{2k}^*\, y_{2k}^*\right]\\[3mm]
    &+M_i^3 M_{\Psi}\,\text{Im}\left[y^2_{2i}\, y_{1k}^*\, y_{2k}^*\right]-3 M_i\,M_{\Psi}^3 \,\text{Im}\left[y^2_{2i}\, y_{1k}^*\, y_{2k}^*\right]-2 M_{\Psi}^4\,\text{Im}\left[y_{1i}\, y_{2i} \,y_{1k}^* \,y_{2k}^*\right]\Bigg\} \\[3mm]
    &-\text{Im}\left[\frac{C_0(M^2_{\Psi},0,M^2_i,0,M^2_k,M^2_{\Psi})}{\pi }\right]\Bigg\{-M_k\, M_{\Psi} \left(-2 \left(M_k^2-M_{\Psi}^2\right)-M_i^4+M_{\Psi}^4\right) \text{Im}\left[y^2_{2i}\, \left(y_{2k}^*\right)^2\right] \\[3mm]
    &+M_k^2\, M_i^4\,\text{Im}\left[y^2_{2i}\, y_{1k}^*\, y_{2k}^*\right]-M_k^2\, M_i^3\, M_{\Psi}\,\text{Im}\left[y_{1i}\, y_{2i}\, y_{1k}^* \,y_{2k}^*\right]+M_k^2\, M_i^3\, M_{\Psi}\,\text{Im}\left[y_{1i}\, y_{2i}\, y_{1k}^*\, y_{2k}^*\right]\\[3mm]
    &-2 M_k^2 \,M_i^2 \,M_{\Psi}^2\,\text{Im}\left[y^2_{2i}\, y_{1k}^*\, y_{2k}^*\right]+M_k^2\, M_i^2\, M_{\Psi}^2\, \text{Im}\left[y^2_{2i}\, y_{1k}^*\, y_{2k}^*\right]\\[3mm]
    &-M_k\, M_i\left(M_i^2+M_{\Psi}^2\right) \left(-M_k^2-M_i^2+2 M_{\Psi}^2\right) \text{Im}\left[y_{1i}\, y_{2i}\, \left(y_{1k}^*\right)^2\right]+2 M_{\Psi}^2 \left(M_{\Psi}^4-M_k^2 M_i^2\right) \text{Im}\left[y^2_{2i}\, y_{1k}^*\, y_{2k}^*\right] \\[3mm]
    &-2 M_k^2\,M_i\,M_{\Psi}^3\,\text{Im}\left[y_{1i}\, y_{2i}\, y_{1k}^*\, y_{2k}^*\right]-M_k\,M_{\Psi} \left(M_i^2-M_{\Psi}^2\right)^2\text{Im}\left[y^2_{2i}\, \left(y_{1k}^*\right)^2\right]-M_i^4\, M_{\Psi}^2\, \text{Im}\left[y^2_{2i}\, y_{1k}^*\, y_{2k}^*\right] \\[3mm]
    &+M_i^4 \,M_{\Psi}^2 \,\text{Im}\left[y^2_{2i}\, y_{1k}^*\, y_{2k}^*\right]-2 M_i^3\, M_{\Psi}^3\,\text{Im}\left[y_{1i}\, y_{2i}\, y_{1k}^*\, y_{2k}^*\right]-M_i^2 M_{\Psi}^4 \,\text{Im}\left[y^2_{2i}\, y_{1k}^*\, y_{2k}^*\right]\\[3mm]
    &+4 M_i\,M_{\Psi}^5\, \text{Im}\left[y_{1i}\, y_{2i}\, y_{1k}^* \,y_{2k}^*\right]+M_{\Psi}^6\, \text{Im}\left[y^2_{2i}\, y_{1k}^* y_{2k}^*\right]\Bigg\}
\end{align*}

\begin{align*}
    &\mathcal{F}_{2a} ~=~  \\
    &-\left(1-\frac{M^2_{\Psi}}{M^2_i}\right)\Theta\left(M^2_i - M^2_{\Psi}\right)\Bigg\{M_k \,M_i \left(M_i^2+M_{\Psi}^2\right) \text{Im}\left[y_{1i}\,y_{2i} \,\left(y_{2k}^*\right)^2\right]+2 M_k \,M_i^2\, M_{\Psi}\, \text{Im}\left[y_{1i}^2 \left(y_{1k}^*\right)^2\right] \\[3mm]
    &+M_i^4 \,\text{Im}\left[(y^2_{1i}\,y_{1k}^*\,y_{2k}^*\right]+M_i^3\, M_{\Psi}\, \text{Im}\left[y_{1i}\,y_{2i}\,y_{1k}^*\,y_{2k}^*\right]-3 M_i^2\, M_{\Psi}^2\, \text{Im}\left[y^2_{1i}\,y_{1k}^*\,y_{2k}^*\right]\\[3mm]
    &-2 M_i \,M_{\Psi}^3 \,\text{Im}\left[y_{1i}\,y_{2i}\,y_{1k}^*\,y_{2k}^*\right]\Bigg\} \\[3mm]
    &+\left(1-\frac{M^2_{k}}{M^2_{\Psi}}\right)\Theta\left( M^2_{\Psi}-M^2_k\right)\Bigg\{M_k\, M_{\Psi} \left(M_i^2+M_{\Psi}^2\right)\text{Im}\left[y_{1i}^2\, \left(y_{1k}^*\right)^2\right]+2 M_k\, M_i\, M_{\Psi}^2\text{Im}\left[y_{1i}\,y_{2i}\,y_{2k}^*\,y_{2k}^*\right]\\[3mm]
    &+M_i^3\, M_{\Psi}\,\text{Im}\left[y_{1i}\,y_{2i}\,y_{1k}^*\,y_{2k}^*\right]-3 M_i\, M_{\Psi}^3\,\text{Im}\left[y_{1i}\,y_{2i}\,y_{1k}^*\,y_{2k}^*\right]-2 M_{\Psi}^4 \,\text{Im}\left[y^2_{1i} \,y_{1k}^*\,y_{2k}^*\right]\Bigg\} \\[3mm]
    &+\text{Im}\left[\frac{C_0(M^2_{\Psi},0,M^2_i,0,M^2_k,M^2_{\Psi})}{\pi }\right]\Bigg\{-M_k^2 M_i^4 \text{Im}\left[y^2_{1i}\, y_{1k}^*\,y_{2k}^*\right]-M_k^2\, M_i^3\, M_{\Psi}\, \text{Im}\left[y_{1i}y_{2i}y_{1k}^*y_{2k}^*\right]\\[3mm]
    &+M_k^2\, M_i^3 \,M_{\Psi}\, \text{Im}\left[y_{1i}\,y_{2i}\,y_{1k}^*\,y_{2k}^*\right]-M_k^2\, M_i^2\, M_{\Psi}^2\, \text{Im}\left[y^2_{1i}\,y_{1k}^*\,y_{2k}^*\right]+2 M_k^2\, M_i^2\, M_{\Psi}^2\,\text{Im}\left[y^2_{1i}\,y_{1k}^*\,y_{2k}^*\right]\\[3mm]
    &+M_k\, M_i\, \left(M_i^2+M_{\Psi}^2\right) \left(-M_k^2-M_i^2-2 M_{\Psi}^2\right) \text{Im}\left[y_{1i}\,y_{2i}\,y_{2k}^*\,y_{2k}^*\right]-2 M_{\Psi}^2 \left(M_{\Psi}^4-M_k^2 M_i^2\right) \text{Im}\left[y^2_{1i}\,y_{1k}^*\,y_{2k}^*\right]\\[3mm]
    &+M_k\, M_{\Psi} \left(-2 M_i^2 \left(M_k^2-M_{\Psi}^2\right)-M_i^4+M_{\Psi}^4\right)\text{Im}\left[y_{1i}^2 \left(y_{1k}^*\right)^2\right]+2 M_k^2\, M_i\, M_{\Psi}^3\,\text{Im}\left[y_{1i}\,y_{2i}\,y_{1k}^*\,y_{2k}^*\right]\\[3mm]
    &+M_k\, M_{\Psi} \left(M_i^2-M_{\Psi}^2\right)^2 \text{Im}\left[y_{1i}^2 \left(y_{2k}^*\right)^2\right]+2 M_i^3\, M_{\Psi}^3 \,\text{Im}\left[y_{1i}\,y_{2i}\,y_{1k}^*\,y_{2k}^*\right]+M_i^2\, M_{\Psi}^4\,\text{Im}\left[y^2_{1i}\,y_{1k}^*\,y_{2k}^*\right]\\[3mm]
    &-4 M_i\, M_{\Psi}^5\, \text{Im}\left[y_{1i}\,y_{2i}\,y_{1k}^*\,y_{2k}^*\right]-M_{\Psi}^6 \text{Im}\left[y^2_{1i}\, y_{1k}^*\,y_{2k}^*\right]\Bigg\}
\end{align*}

\begin{align*}
    &\mathcal{F}_{2b} ~=~  \\
    &\left(1-\frac{M^2_{\Psi}}{M^2_i}\right)\Theta\left(M^2_i - M^2_{\Psi}\right)\Bigg\{-2 M_k^2\, M_i\, M_{\Psi}\,\text{Im}\left[y^2_{2i}\, \left(y_{2k}^*\right)^2\right]+M_k\, M_i \left(M_i^2+M_{\Psi}^2\right) \,\text{Im}\left[\left(y_{1k}^*\right)^2 y_{1i}\,y_{2i}\right]\\[3mm]
    &-M_i^4 \,\text{Im}\left[y^2_{2i}\,y_{1k}^*\,y_{2k}^*\right]-M_i^3\, M_{\Psi}\,\text{Im}\left[y_{1i}\,y_{2i}\,y_{1k}^*\,y_{2k}^*\right]+3 M_i^2\, M_{\Psi}^2\,\text{Im}\left[y^2_{2i}\,y_{1k}^*\,y_{2k}^*\right]\\[3mm]
    &+2 M_i\, M_{\Psi}^3\,\text{Im}\left[y_{1i}\,y_{2i}\,y_{1k}^*\,y_{2k}^*\right]\Bigg\} \\[3mm]
    &-\left(1-\frac{M^2_{k}}{M^2_{\Psi}}\right)\Theta\left( M^2_{\Psi}-M^2_k\right)\Bigg\{-M_k\, M_{\Psi} \left(M_i^2+M_{\Psi}^2\right) \text{Im}\left[y^2_{2i}\,y_{2k}^*\,y_{2k}^*\right]-2 M_k\, M_i\, M_{\Psi}^2\,\text{Im}\left[\left(y_{1k}^*\right)^2 \,y_{1i}\,y_{2i}\right] \\[3mm]
    &-M_i^3\, M_{\Psi}\, \text{Im}\left[y_{1i}\,y_{2i}\,y_{1k}^*\,y_{2k}^*\right]+3 M_i\, M_{\Psi}^3\,\text{Im}\left[y_{1i}\,y_{2i}\,y_{1k}^*\,y_{2k}^*\right]+2 M_{\Psi}^4\,\text{Im}\left[y^2_{2i}\,y_{1k}^*\,y_{2k}^*\right]\Bigg\} \\[3mm]
    &-\text{Im}\left[\frac{C_0(M^2_{\Psi},0,M^2_i,0,M^2_k,M^2_{\Psi})}{\pi }\right]\Bigg\{-M_k M_{\Psi} \left(-2 \left(M_k^2-M_{\Psi}^2\right)-M_i^4+M_{\Psi}^4\right) \text{Im}\left[y^2_{2i} \left(y_{2k}^*\right)^2\right]\\[3mm]
    &+M_k^2\, M_i^4\,\text{Im}\left[y^2_{2i}\,y_{1k}^*\,y_{2k}^*\right]- M_k^2\, M_i^2\, M_{\Psi}^2\, \text{Im}\left[y^2_{2i}\,y_{1k}^*\,y_{2k}^*\right]\\[3mm]
    &-M_k\, M_i \left(M_i^2+M_{\Psi}^2\right) \left(-M_k^2-M_i^2+2 M_{\Psi}^2\right)\,\text{Im}\left[y_{1i}\,y_{2i}\, \left(y_{1k}^*\right)^2\right]+2 M_{\Psi}^2\, \left(M_{\Psi}^4-M_k^2 M_i^2\right) \text{Im}\left[y^2_{2i}\,y_{1k}^*\,y_{2k}^*\right]\\[3mm]
    &-2 M_k^2\, M_i\, M_{\Psi}^3\, \text{Im}\left[y_{1i}\,y_{2i}\,y_{1k}^*\,y_{2k}^*\right]
    -M_k\, M_{\Psi} \left(M_i^2-M_{\Psi}^2\right)^2 \text{Im}\left[y^2_{2i}\,\left(y_{1k}^*\right)^2\right]-M_i^4\, M_{\Psi}^2\, \text{Im}\left[y^2_{2i}\,y_{1k}^*\,y_{2k}^*\right]\\[3mm]
    &+M_i^4\, M_{\Psi}^2\,\text{Im}\left[y^2_{2i}\,y_{1k}^*\,y_{2k}^*\right]-2 M_i^3\, M_{\Psi}^3\,\text{Im}\left[y_{1i}\,y_{2i}\,y_{1k}^*\,y_{2k}^*\right]-M_i^2\, M_{\Psi}^4\,\text{Im}\left[y^2_{2i}\,y_{1k}^*\,y_{2k}^*\right]\\
    &+4 M_i\, M_{\Psi}^5\, \text{Im}\left[y_{1i}\,y_{2i}\,y_{1k}^*\,y_{2k}^*\right]+M_{\Psi}^6\, \text{Im}\left[y^2_{2i}\,y_{1k}^*\,y_{2k}^*\right]\Bigg\}
\end{align*}

\begin{align}
    &\text{V}^{(i)}_3 ~=~ 
    \frac{1}{128\pi^2} \frac{M_i\left(1-\frac{M^2_{\eta}}{M_i^2}\right)^2}{\Gamma^i_{\text{tot.}}}\displaystyle\sum_{j,k = 1}^3 \text{Im}\left[g^j_{F_i}\,y^{\text{SM}}_{jk}\,g^k_R\,y_{1i}^{\ast}\right]\Bigg\{\left(1-\frac{M^2_{\Psi}}{M^2_{\eta}}\right)\,\Theta\left(M^2_{\eta}-M^2_{\Psi}\right) \nonumber \\[3mm] 
    & \qquad \qquad \qquad \qquad +\frac{M^2_{i}}{M^2_{i} - M^2_{\eta}}\bigg[\left(1-\frac{M^2_{\Psi}}{M^2_{i}}\right)\,\Theta\left(M^2_{i}-M^2_{\Psi}\right) - \left(1-\frac{M^2_{\Psi}}{M^2_{\eta}}\right)\,\Theta\left(M^2_{\eta}-M^2_{\Psi}\right)\bigg]\Bigg\} \nonumber \\[3mm]
    & \qquad  -\frac{1}{128\pi^2} \frac{M_i\left(1-\frac{M^2_{\eta}}{M_i^2}\right)^2}{\Gamma^i_{\text{tot.}}}\displaystyle\sum_{j,k = 1}^3\text{Im}\left[g^j_{F_i}\,y^{\text{SM}}_{jk}\,g^k_R\,y_{2i}^{\ast}\right]\frac{M_{i}\,M_{\Psi}}{M^2_{i} - M^2_{\eta}}\Bigg\{\left(1-\frac{M^2_{\Psi}}{M^2_{i}}\right)\,\Theta\left(M^2_{i}-M^2_{\Psi}\right) \nonumber \\[3mm]
    &\qquad \qquad \qquad \qquad- \left(1-\frac{M^2_{\Psi}}{M^2_{\eta}}\right)\,\Theta\left(M^2_{\eta}-M^2_{\Psi}\right)\Bigg\}
\end{align}

\begin{align}
    &\text{V}^{(i)}_4 ~=~ -\frac{1}{128\pi^2}\frac{M_i\left(1-\frac{M^2_{\Psi}}{M^2_i}\right)}{\Gamma^i_{\text{tot.}}} 
    \displaystyle\sum_{j,k = 1}^3 \text{Im}\left[(g^j_{F_i})^{\ast}\,(y^{\text{SM}}_{jk})^{\ast}\,(g^k_R)^{\ast}\,y_{1i}\right] \Bigg\{\left(1 + \frac{M^2_{\Psi}}{M^2_i}\right)\mathcal{F}_3\left(M^2_{i},M^2_{\Psi},M^2_{\eta}\right) \nonumber \\
    &\qquad \qquad \qquad -2\frac{M_{\Psi}}{M_i}\mathcal{F}_4\left(M^2_{i},M^2_{\Psi},M^2_{\eta}\right)\Bigg\} \nonumber \\[3mm]
    &-\frac{1}{128\pi^2}\frac{M_i\left(1-\frac{M^2_{\Psi}}{M^2_i}\right)}{\Gamma^i_{\text{tot.}}} \displaystyle\sum_{j,k = 1}^3 \text{Im}\left[(g^j_{F_i})^{\ast}\,(y^{\text{SM}}_{jk})^{\ast}\,(g^k_R)^{\ast}\,y_{2i}\right] \Bigg\{\left(1 + \frac{M^2_{\Psi}}{M^2_i}\right)\mathcal{F}_4\left(M^2_{i},M^2_{\Psi},M^2_{\eta}\right)\nonumber \\
    &\qquad \qquad \qquad-2\frac{M_{\Psi}}{M_i}\mathcal{F}_3\left(M^2_{i},M^2_{\Psi},M^2_{\eta}\right)\Bigg\}
\end{align}
where 
\begin{align*}
    \mathcal{F}_3\left(M^2_{i},M^2_{\Psi},M^2_{\eta}\right) ~=~ \frac{1}{M^2_i - M^2_{\Psi}}\Bigg[M^2_{\Psi}\,&\left(1-\frac{M^2_{\eta}}{M^2_{\Psi}}\right)\,\Theta(M^2_{\Psi}-M^2_{\eta}) 
    -M^2_{i}\,\left(1-\frac{M^2_{\eta}}{M^2_{i}}\right)\,\Theta(M^2_{i}-M^2_{\eta}) \\[3mm]
    &-2 M^2_i\,(M^2_{\Psi}-M^2_{\eta})\,\text{Im}\left[\frac{C_0(M^2_{\psi},0,M^2_i,M^2_{\eta},0,0)}{\pi}\right]\Bigg] 
\end{align*}

\begin{align*}
    \mathcal{F}_4\left(M^2_{i},M^2_{\Psi},M^2_{\eta}\right) ~=~ \frac{M_i\,M_{\Psi}}{M^2_i - M^2_{\Psi}}\Bigg[\left(1-\frac{M^2_{\eta}}{M^2_{\Psi}}\right)\,\Theta(M^2_{\Psi}-M^2_{\eta})-\left(1-\frac{M^2_{\eta}}{M^2_{i}}\right)\,\Theta(M^2_{i}-M^2_{\eta})\Bigg]
\end{align*}
In $\text{V}^{(3,4)}$, $y^{\text{SM}}$ refers to the Standard Model lepton-Higgs Yukawa couplings.

\begin{align}
    &\text{V}^{(i)}_5 ~=~ 
    \frac{1}{128\pi^2} \frac{M_i\left(1-\frac{M^2_{\eta}}{M_i^2}\right)^2}{\Gamma^i_{\text{tot.}}}\displaystyle\sum_{j=1}^3\text{Im}\left[\alpha \,g^j_{F_i}\,(g^j_{\Psi})^{\ast}\,y_{1i}^{\ast}\right]\,M_{\Psi}\,\text{Im}\left[\frac{C_0(0,M^2_{\eta},M^2_i,M^2_{\Psi},M^2_S,0)}{\pi}\right] \nonumber \\[3mm]
    & -\frac{1}{128\pi^2} \frac{M_i\left(1-\frac{M^2_{\eta}}{M_i^2}\right)^2}{\Gamma^i_{\text{tot.}}}\displaystyle\sum_{j=1}^3\text{Im}\left[\alpha \,g^j_{F_i}\,(g^j_{\Psi})^{\ast}\,y_{2i}^{\ast}\right]\,M_i\,\Bigg\{\frac{1}{M^2_i-M^2_{\eta}}\left(1-\frac{M^2_S}{M^2_{\eta}}\right)\Theta\left(M^2_{\eta}-M^2_S\right) \nonumber \\[3mm]
    &-\frac{1}{M^2_i-M^2_{\eta}}\left(1-\frac{M^2_{\Psi}}{M^2_{i}}\right)\Theta\left(M^2_{i}-M^2_{\Psi}\right) + \frac{M^2_{\Psi}-M^2_S}{M^2_i-M^2_{\eta}}\,\text{Im}\left[\frac{C_0(0,M^2_{\eta},M^2_i,M^2_{\Psi},M^2_S,0)}{\pi}\right]\Bigg\}
\end{align}

\begin{align}
    &\text{V}^{(i)}_6 ~=~ -\frac{1}{128\pi^2}\frac{M_i\left(1-\frac{M^2_{\Psi}}{M^2_i}\right)}{\Gamma^i_{\text{tot.}}} \displaystyle\sum_{j= 1}^3 \text{Im}\left[\alpha^{\ast}\,(g^j_{F_i})^{\ast}\,g^j_{\Psi}\,y_{1i}\right] \Bigg\{\left(1+\frac{M^2_{\Psi}}{M^2_i}\right) \mathcal{F}_5\left(M_{i},M_{\Psi},M_{\eta},M_{S}\right) \nonumber \\[3mm]
    &\qquad \qquad \qquad -2\frac{M_{\Psi}}{M_{i}} \mathcal{F}_6\left(M_{i},M_{\Psi},M_{\eta},M_{S}\right)\Bigg\} \nonumber \\[3mm]
    &-\frac{1}{128\pi^2}\frac{M_i\left(1-\frac{M^2_{\Psi}}{M^2_i}\right)}{\Gamma^i_{\text{tot.}}} \displaystyle\sum_{j= 1}^3 \text{Im}\left[\alpha^{\ast}\,(g^j_{F_i})^{\ast}\,g^j_{\Psi}\,y_{2i}\right] \Bigg\{\left(1+\frac{M^2_{\Psi}}{M^2_i}\right) \mathcal{F}_6\left(M^2_{i},M^2_{\Psi},M^2_{\eta},M^2_{S}\right) \nonumber \\[3mm]
    &\qquad \qquad \qquad -2\frac{M_{\Psi}}{M_{i}} \mathcal{F}_5\left(M^2_{i},M^2_{\Psi},M^2_{\eta},M^2_{S}\right)\Bigg\}
\end{align}
where 
\begin{align*}
    &\mathcal{F}_5\left(M_{i},M_{\Psi},M_{\eta},M_{S}\right) ~=~ \\[3mm] 
    &\frac{M_{\Psi}}{(M^2_i-M^2_{\Psi})^2}\Bigg[(M^2_i+M^2_{\Psi})\,\left(1-\frac{M^2_S}{M^2_{\Psi}}\right)\Theta(M^2_{\Psi}-M^2_S)  -2M^2_i\left(1-\frac{M^2_{\eta}}{M^2_{i}}\right)\Theta(M^2_{i}-M^2_{\eta}) \\[3mm]
    & \qquad \qquad \qquad \qquad +(M^4_i-M^2_i(M^2_{\Psi}+M^2_{\eta}-2M^2_S)-M^2_{\Psi}M^2_{\eta})\,\text{Im}\left[\frac{C_0(M^2_{\psi},0,M^2_i,0,M^2_S,M^2_{\eta})}{\pi}\right]\Bigg]
\end{align*}

\begin{align*}
    &\mathcal{F}_6\left(M_{i},M_{\Psi},M_{\eta},M_{S}\right) ~=~\\[3mm] 
    &\frac{M_i}{(M^2_i-M^2_{\Psi})^2}\Bigg[2M^2_{\Psi}\,\left(1-\frac{M^2_S}{M^2_{\Psi}}\right)\Theta(M^2_{\Psi}-M^2_S) - (M^2_i+M^2_{\Psi})\,\left(1-\frac{M^2_{\eta}}{M^2_{i}}\right)\Theta(M^2_{i}-M^2_{\eta}) \\[3mm]
    &\qquad \qquad \qquad \qquad + (M^4_{\Psi}-M^2_{\Psi}(M^2_{i}+M^2_{S}-2M^2_{\eta})-M^2_{i}M^2_{S})\,\text{Im}\left[\frac{C_0(M^2_{\psi},0,M^2_i,0,M^2_S,M^2_{\eta})}{\pi}\right]\Bigg]
\end{align*}
We have verified that the expressions $\text{WF}^{(i)}_1$ and $\text{V}^{(i)}_1$ reproduce the known results \cite{Covi:1996wh} in the limiting case of $M_{\eta} \rightarrow 0$. 

% ===========================================================================

\bibliographystyle{JHEP}
\bibliography{paper}

% ===========================================================================
\end{document}